\documentclass[a4paper,fleqn]{cas-dc}

\usepackage[numbers,sort&compress]{natbib}
\usepackage{amsmath,amssymb,amsfonts}
\usepackage{algorithm}
\usepackage{algorithmic}
\usepackage{graphicx}
\usepackage{textcomp}
\usepackage[table]{xcolor}
\usepackage{multirow}
\usepackage{verbatim}
\usepackage{makecell}
\usepackage{stfloats}
\usepackage{bbding}
\usepackage{framed}
\usepackage{latexsym}
\usepackage{url}
\usepackage{pifont}
\usepackage{arydshln}
\usepackage{bm}
\usepackage{booktabs}
\usepackage{multicol}
\usepackage{threeparttable}
\usepackage{wrapfig}
\usepackage{ulem}
\usepackage{diagbox}

\begin{document}
\let\WriteBookmarks\relax
\def\floatpagepagefraction{1}
\def\textpagefraction{.001}

\shorttitle{KHiM-Mamba: Injecting Pathology Knowledge into Mamba via Hidden-State Modulation for Whole Slide Image Analysis}
\shortauthors{Q. Zhang et al.}

\title[mode=title]{KHiM-Mamba: Injecting Pathology Knowledge into Mamba via Hidden-State Modulation for Whole Slide Image Analysis}
\author[label1]{Qixiang Zhang}[orcid=0009-0002-3158-9471]
\author[label1]{Yi Li}[orcid=0000-0002-7840-2611]
\author[label1]{Tianqi Xiang}[orcid=0000-0002-5550-1721]
\author[label1]{Haonan Wang}[orcid=0000-0001-8241-7982]
\author[label2]{Mengjiao Wei}
\author[label2]{Bo Xu}[orcid=0000-0001-8693-3060]
\author[label1]{Xiaomeng Li}[orcid=0000-0003-1105-8083]
\cormark[1]
\ead{Corresponding author (eexmli@ust.hk)}

\affiliation[label1]{organization={Department of Electronic and Computer Engineering, The Hong Kong University of Science and Technology},
            city={Hong Kong},
            country={China}}
\affiliation[label2]{organization={Center for Intelligent Oncology, Chongqing University Cancer Hospital and Chongqing University School of Medicine, and Chongqing Key Laboratory of Intelligent Oncology for Breast Cancer},
            city={Chongqing},
            postcode={400030}, 
            country={China}}

\begin{abstract}
Whole-slide image analysis is commonly formulated as multiple instance learning (MIL), where instance features are contextually updated and aggregated into a slide representation—a process we term slide encoding dynamics. Recently, selective state-space models (SSM) have emerged as promising MIL architectures due to their long-sequence modeling capability and linear complexity. However, existing SSM-based MIL methods rely solely on visual features during MIL. Meanwhile, in large-scale WSIs, where sparse diagnostically decisive regions are surrounded by abundant irrelevant information, such purely vision-driven selective dynamics can misallocate state updates and readouts, causing the evolving SSM state to accumulate task-irrelevant evidence and dilute critical diagnostic cues over long scan trajectories. In this work, we propose the \textbf{K}nowledge-Aware \textbf{Hi}dden-State \textbf{M}odulation architecture (KHiM-Mamba), which innovatively regulates Mamba's core selective state-space mechanism with explicit knowledge priors, steering slide encoding dynamics toward diagnostically meaningful evidence accumulation. Specifically, we redesign the original SSM layer to perform knowledge modulation operations during the evolution of hidden states, thereby guiding what visual evidence is accumulated and retrieved from the hidden state at each encoding step. Furthermore, we additionally introduce a local-adaptive vocabulary retrieval module that uses large language models to assign each patch fine-grained, tissue-specific semantic descriptions, enabling precise modulation across diverse tasks. Experiments on 11 public benchmarks across 4 tasks show that KHiM-Mamba consistently achieves state-of-the-art performance.
\end{abstract}


\maketitle

\definecolor{softred}{RGB}{180,90,90}
\definecolor{softgreen}{RGB}{90,140,100}

\section{Introduction}
\label{sec:introduction}
\begin{figure}[pos=t]
    \centering
    \includegraphics[width=0.48\textwidth]{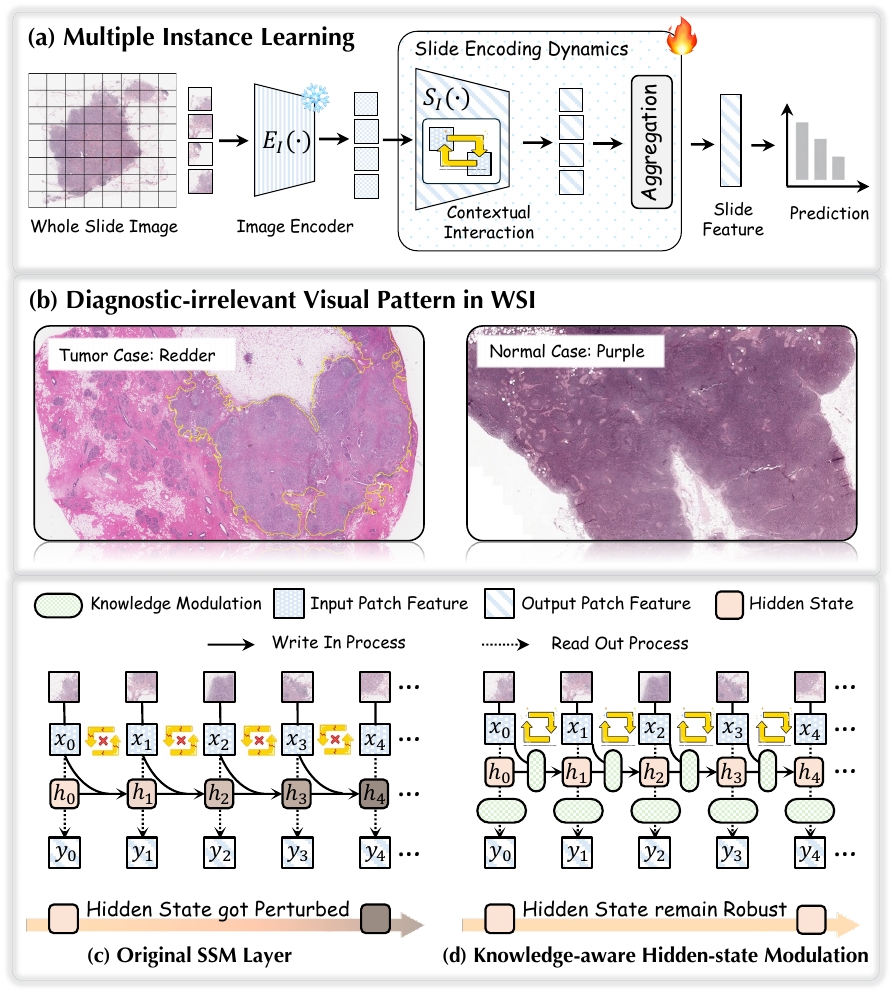}
    \caption{(a) MIL pipeline: a WSI is split into patches, encoded as patch features, and processed by slide encoding dynamics (\textit{i.e.}, contextual interaction and aggregation). (b) Example of diagnostically irrelevant visual patterns in WSIs: under a certain stain, tumor slides appear redder while normal slides appear purpler. (c)(d) Comparison of SSM layer in prior Mamba-based MIL methods and ours.} \label{fig:intro_1}
    \vspace{-0.5cm}
\end{figure}

Computational pathology (CPath) has advanced rapidly with the development of artificial intelligence for pathological analysis~\cite{mil_survey,qi2025depth,cpath_review_1}. One core element in CPath is the whole slide image (WSI), a digital scan of a pathology slide. Uniquely, WSIs can reach gigapixel resolution (\textit{e.g.}, 80,000$\times$80,000 pixels). Such massive scale has motivated the development of multi-instance learning (MIL)~\cite{mil_survey,cpath_review_1}. In a typical MIL pipeline, a pre-trained patch encoder~\cite{uni,conch,plip} first extracts a feature for each small patch of the WSI independently, and a slide encoder then models inter-patch dependencies to facilitate contextual interaction between patch features and aggregate them into a slide-level representation for downstream prediction. We term this process---encompassing inter-patch contextual interaction and representation formation---\textit{slide encoding dynamics} (Fig.~\ref{fig:intro_1}a). It is central to WSI analysis since diagnostic evidence is often distributed across distant tissue regions and should be accumulated from multiple patches.

\begin{figure*}[pos=t]
    \centering
    \includegraphics[width=\textwidth]{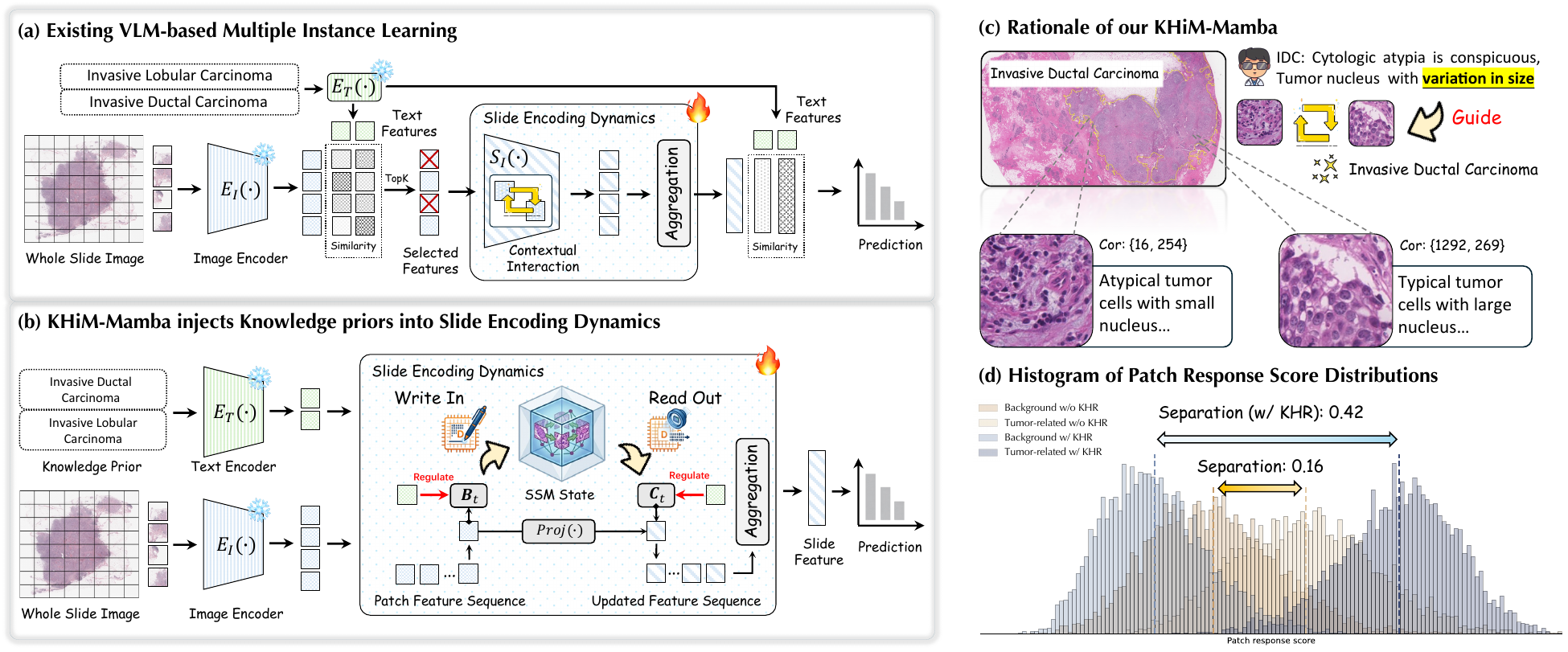}
    \caption{(a) Existing VLM-MIL methods typically inject language priors either before slide encoding~\cite{focus,top} to guide patch selection via image-text similarity, or after slide encoding as auxiliary supervision~\cite{vilamil}. (b) KHiM-Mamba injects knowledge priors into the core slide encoding dynamics by modulating two key matrix: input-to-state $B_t$, which governs how each patch’s information is written into the hidden state, and state-to-output $C_t$, which governs how accumulated context is read out to enrich patch representations. 
    (c) An illustrative example of language-guided contextual interaction across patches. In invasive ductal carcinoma, \texttt{``conspicuous cytologic atypia and variation in nuclear size''} guides the model to relate distinct patches with complementary nuclear morphologies, thereby grounding the slide-level representation in pathology-relevant evidence.
    (d) Histograms of patch response scores for background and tumor-related patches, with and without our Knowledge-aware Hidden-state Modulation (KHiM). KHiM markedly reduces the overlap between the two distributions, indicating stronger patch selectivity and clearer semantic separation; More quantitative and qualitative results to support this statement can be found in Tab.~\ref{ablation} and Fig.~\ref{fig:vis} respectively} \label{fig:intro_2}
    \vspace{-0.5cm}
\end{figure*}

Various architectures (\textit{e.g.}, attention~\cite{abmil,transmil_1,dsmil,dtfdmil} and graph~\cite{GraphWSI}) have been explored as slide encoders to improve \textit{slide encoding dynamics}. More recently, the Selective State Space Model (Mamba)~\cite{mamba} has emerged as a compelling alternative for WSI analysis (\textit{e.g.}, PAM~\cite{PAM}, GMMamba~\cite{gmmamba}, M3Mamba~\cite{m3mamba}), achieving state-of-the-art performance. Existing Mamba-based MIL propose to scan WSIs into long sequences of patches via tailored scanning strategies (\textit{e.g.}, PAM's hierarchical scanning~\cite{PAM}, GMMamba's grouping scanning~\cite{gmmamba}), and then rely on Mamba's native state space model (SSM) layer to perform contextual interaction during slide encoding dynamics. However, we identify a fundamental limitation in existing Mamba-based MIL methods. The inherited SSM layer used by native Mamba could be easily affected by abundant diagnostically irrelevant variations in WSIs, causing irrelevant cues to bias contextual interactions. For example, due to stain bias in the CAMELYON dataset~\cite{camelyon16}, tumor slides may appear redder, whereas normal slides appear more purple (see Fig.~\ref{fig:intro_1}b). Such specific but diagnostic-irrelevant visual patterns may involve irrelevant information into SSM's evolving hidden state. Consequently, as a crucial intermediate, the perturbed hidden state can further bias the contextual interactions across patches (Fig.~\ref{fig:intro_1}c), thereby hindering the effective accumulation of diagnostically relevant information from patch features.

To address this limitation, we draw inspiration from prior VLM-MIL studies~\cite{vilamil,moc,top,focus,muse}, which have highlighted the value of language-derived diagnostic knowledge for WSI modeling. Rather than solely relying on vulnerable visual features from WSIs, we introduce \textbf{K}nowledge-aware \textbf{Hi}dden-state \textbf{M}odulation (KHiM-Mamba), which injects knowledge priors into the SSM layer to steer Mamba-based slide encoding toward diagnostically relevant evidence (see Fig.\ref{fig:intro_1}cd). \textit{Crucially, our innovation lies not in the use of language per se, but in how language participates in the slide encoding process.} Existing VLM-MIL methods typically employ language as an external add-on at the periphery of slide encoding, either to select patches before encoding or to provide auxiliary supervision afterward (Fig.\ref{fig:intro_2}a). Since language does not intervene in the core slide encoding dynamics, contextual interactions among WSI patches remain governed solely by visual features, leaving the aforementioned limitation unresolved. In contrast, KHiM-Mamba redesigns the information flow within the SSM layer (see Fig.~\ref{fig:intro_1}d), allowing knowledge priors to directly intervene in the transition of the evolving hidden state within the slide encoding dynamics (Fig.~\ref{fig:intro_2}b). The underlying rationale of this design is that explicit diagnostic knowledge conveyed by language can guide contextual interactions across patches toward diagnostically relevant relations among distinct WSI regions, as exemplified by the case shown in Fig.~\ref{fig:intro_2}c. To realize this idea, we reformulate the SSM layer with two novel knowledge modulation schemes: write-in modulation of the input-to-state matrix, which guides what visual evidence is incorporated into the evolving state, and read-out modulation of the state-to-output matrix, which guides what accumulated contextual evidence is retrieved to enrich patch representations. Through these mechanisms, each patch interacts with the shared context under explicit semantic guidance, promoting the accumulation of diagnostically meaningful evidence (see Fig.\ref{fig:intro_2}d).

The subsequent step is to construct effective language priors for the proposed modulation. In order to provide effective language priors for such modulation, we design a local-adaptive vocabulary retrieval module via large language models (LLMs), which identifies task-relevant tissue concepts and dynamically assigns each patch fine-grained, tissue-specific language based on its local tissue type, yielding modulation signals that are both semantically precise and robust across diverse downstream tasks.

The main contributions are summarized as follows:
\begin{itemize}
    \item We identify a fundamental limitation of existing Mamba-based MIL methods: purely vision-driven state transitions may accumulate diagnostically irrelevant evidence and dilute critical cues over long WSI sequences. Meanwhile, existing VLM-MIL methods employ language only at the periphery of slide encoding, leaving this limitation fundamentally unresolved.
    \item We propose KHiM-Mamba, which redesigns the SSM layer by modulating the hidden states evolution via language-derived knowledge. To our knowledge, this is the first work to reshape Mamba's internal model dynamics for WSI analysis, rather than using it as a ready-made backbone.
    \item We propose local-adaptive vocabulary retrieval module that uses LLMs to dynamically assign each patch fine-grained language, offering semantically precise modulation signals across diverse downstream tasks.
    \item Extensive experiments on 11 public benchmarks across cancer subtyping (\textbf{\textcolor{softgreen}{+2.07\%}}) and survival prediction (\textbf{\textcolor{softgreen}{+2.31\%}}) demonstrate that KHiM-Mamba consistently outperforms state-of-the-art methods, with strong generalization under multi-center domain adaptation (\textbf{\textcolor{softgreen}{+2.27\%}}) and few-shot scenarios (\textbf{\textcolor{softgreen}{+7.97\%}}).
\end{itemize}

\section{Related Works}
\label{sec:related_work}
\subsection{Multi-instance Learning in WSI Analysis}
Multiple instance learning (MIL) is the standard framework for whole-slide image (WSI) analysis in computational pathology~\cite{mil_survey,cpath_review_1}. Typical MIL pipelines consist of patch extraction, patch encoding with a pre-trained backbone, and \textit{slide encoding dynamics} for slide-level prediction. Early methods use simple max or mean pooling, while later works introduce learnable aggregation modules, including attention-based pooling (ABMIL~\cite{abmil}), multi-branch design (CLAM~\cite{clam}), dual-stream aggregation (DSMIL~\cite{dsmil}), transformer-based modeling (TransMIL~\cite{transmil_1}), pseudo-bag optimization (DtfdMIL~\cite{dtfdmil}), and deconfounding (IBMIL~\cite{IBMIL}). Despite their effectiveness, these methods primarily improve the aggregation of visual instance features, leaving slide-level representations formed purely from visual cues. In contrast, our method incorporates language priors into the \textit{slide encoding dynamics} process, enabling distributed diagnostic evidence to be propagated and integrated under explicit semantic guidance. To this end, we introduce KHiM-Mamba that dynamically controls information propagation during slide encoding.

\subsection{Selective State Space Model (Mamba)}
State space models (SSMs)~\cite{s4_1,mamba} have recently emerged as a scalable alternative to conventional sequence architectures for long-context modelling. By representing sequences with hidden states and transition dynamics rather than explicit pairwise interactions, SSMs offer linear complexity and avoid the quadratic cost of Transformers. Two representative examples are S4~\cite{s4_1} and S6 (Mamba)~\cite{mamba}, which have shown strong performance across both vision~\cite{vmamba} and language~\cite{dao2024transformers} tasks. In computational pathology, S4MIL demonstrates that SSM-based models can effectively process gigapixel WSIs by capturing long-range dependencies among patch sequences. Subsequent Mamba-based methods further improve robustness and efficiency: MambaMIL~\cite{mambamil,mambamil+} reduces the impact of noisy instances via sequence reordering, M3amba lowers scanning cost using a memory bank of representative instance clusters, GMMamba~\cite{gmmamba} skips redundant instances through attention-based grouping, and 2DMamba~\cite{2dmamba} preserves local spatial structure with a two-dimensional scanning scheme. Existing Mamba-based MIL methods mainly use Mamba as a ready-made backbone and focus on patch scanning strategies for 1D sequence construction, while leaving its selective state transition unchanged. As a result, slide encoding remains a purely visual process. In contrast, our method adopts a naive bi-directional Mamba with vanilla scanning~\cite{vmamba}, and focus on fundamental backbone-level improvements to the core selective state transitions. Our KHiM-Mamba injects language priors into the selective state transition itself, turning Mamba into a semantically guided representation formation mechanism. This design is orthogonal to existing scanning strategies and can be readily combined with previous methods.

\subsection{Vision-language Multiple Instance Learning}
General-purpose vision-language models, such as CLIP, have demonstrated strong transferability across a wide range of visual tasks, inspiring pathology-specific foundation models including PLIP~\cite{plip}, CONCH~\cite{conch}, and MUSK~\cite{musk}. Building on these advances, recent studies have combined vision-language models with MIL for WSI analysis, aiming to exploit their few-shot and zero-shot capabilities, as exemplified by TOP~\cite{top}, ViLa-MIL~\cite{vilamil}, FOCUS~\cite{focus}, and MUSE~\cite{muse}. Despite their promise, existing VLM-MIL methods typically use language in a relatively global or peripheral manner, e.g., through class-level prompts, patch selection, or auxiliary supervision, rather than integrating it into the instance-level encoding process itself (see Fig.~\ref{fig:intro_2}a). As a result, language semantics have limited influence on how patch features interact and how diagnostic evidence is accumulated during slide encoding. In contrast, our method treats language as a dynamic diagnostic prior and injects language-conditioned modulation into the core \textit{slide encoding dynamics}, enabling semantically guided patch interaction and evidence integration.

\begin{figure*}[pos=t]
    \centering
    \includegraphics[width=\textwidth]{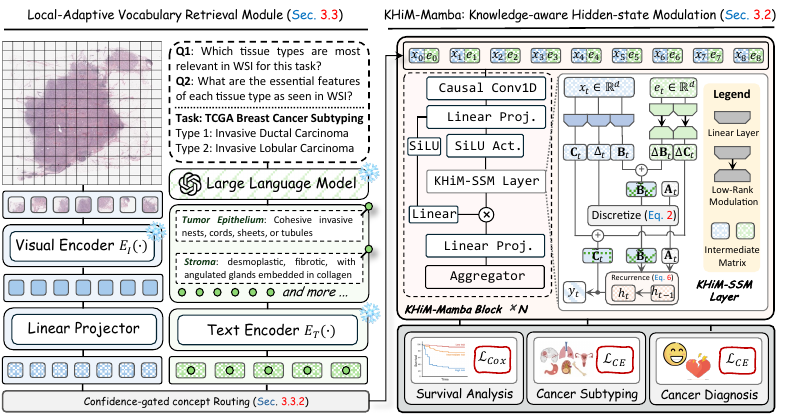}
    \caption{Overview of \textbf{\textit{KHiM-Mamba}}. Our framework consists of two main blocks: Local-adaptive vocabulary retrieval module (Sec.~\ref{sec:vocab}) provides precise language priors across diverse downstream tasks; and \textbf{K}nowledge-aware \textbf{Hi}dden-state \textbf{M}odulation (KHiM-Mamba, Sec.~\ref{sec:KHiM-Mamba}) which leverages language priors to modulate the Mamba-based slide encoding dynamics.}
    \label{fig_method}
    \vspace{-0.3cm}
\end{figure*}

\section{Methodology}
\subsection{Preliminary}
\label{sec:pre}
Selective State Space Models (SSMs), such as Mamba~\cite{s4_1,mamba}, provide an efficient framework for long-sequence modeling and have recently shown strong potential in computational pathology~\cite{S4MIL,mambamil,PAM}. Mamba builds upon the continuous-time SSM:
\begin{equation}
    h'(t)=Ah(t)+Bx(t), \quad y(t)=Ch(t),
\end{equation}
where \(h(t)\) denotes the latent state, and \(A\), \(B\), \(C\) are the state evolution, input projection, and output projection matrices, respectively. To process discrete sequences, the continuous system is discretized with an input-dependent time-scale parameter \(\Delta_t\), yielding the recurrent form:
\begin{equation}
    h_t = \bar{A}_t \, h_{t-1} + \Delta_t \mathbf{B}_t \, x_t, \quad y_t = \mathbf{C}_t \, h_t, \quad \bar{A}_t = \exp(\Delta_t A), 
\label{eq:recurrent}
\end{equation}

where \(h_t\) serves as an evolving contextual memory. Mamba's \emph{selective} mechanism makes \(\Delta_t\), \(\mathbf{B}_t\), and \(\mathbf{C}_t\) input-dependent: \(\mathbf{B}_t\) governs how the current input is \emph{written} into \(h_t\), while \(\mathbf{C}_t\) governs how context is \emph{read} from \(h_t\). Our proposed Knowledge-aware Hidden-state Modulation (KHiM-Mamba) modulates both projections with language-derived knowledge priors to reshape this write--read dynamics, as detailed in Sec.~\ref{sec:KHiM-Mamba}.

\subsection{Knowledge-aware Hidden-state Modulation}
\label{sec:KHiM-Mamba}
Based on the above formulation, in this section, we describe how to leverage language priors to modulate Mamba to guide diagnosis-aware evidence propagation across patch sequences. Let \(X=\{x_t\}_{t=1}^{T}\) denote the patch sequence of a whole slide image with $T$ patches, where \(x_t \in \mathbb{R}^{d}\) is the visual feature of the $t^{th}$ patch. Our goal is to modulate the selective dynamics of Mamba with language guidance, so that information flow across patches is steered by diagnosis-related semantics rather than relying solely on visual appearance.

As established in the Sec.~\ref{sec:pre}, \(\mathbf{B}_t\) and \(\mathbf{C}_t\) are the input-dependent projections that respectively control how information is written into and read from the hidden memory \(h_t\). We choose to modulate these two projections rather than the time-scale parameter \(\Delta_t\) or the state matrix \(A\), because conditioning \(\mathbf{B}_t\) and \(\mathbf{C}_t\) does not alter the temporal dynamics or decay behaviour of the SSM; instead, it reshapes the information injection and readout channels in a manner analogous to adapting linear projections, leading to more stable optimization.

Concretely, we introduce a low-rank language modulation mechanism that adds language-conditioned increments to both \(\mathbf{B}_t\) and \(\mathbf{C}_t\). Let \(\mathbf{e}_t \in \mathbb{R}^{d}\) denote a language embedding that encodes diagnostic semantics relevant to patch \(x_t\); the construction of \(\mathbf{e}_t\) is detailed in Sec.~\ref{sec:vocab}. A shared projection maps \(\mathbf{e}_t\) to a pair of low-rank coefficient vectors:
\begin{equation}
    [\boldsymbol{\alpha}^B_t ; \boldsymbol{\alpha}^C_t] = \mathrm{MLP}(\mathbf{e}_t) \in \mathbb{R}^{2r},
\end{equation}
where \(r \ll N\) is the modulation rank, \(N\) is the SSM state dimension, and \([\cdot\,;\cdot]\) denotes concatenation. The coefficients are then composed with learnable basis matrices to form the modulation increments:
\begin{equation}
    \Delta B_t = \boldsymbol{\alpha}^B_t \, \mathbf{W}_B, \quad \Delta C_t = \boldsymbol{\alpha}^C_t \, \mathbf{W}_C,
\end{equation}
where \(\mathbf{W}_B, \mathbf{W}_C \in \mathbb{R}^{r \times N}\) are learnable. The modulated projections are obtained by residual addition:
\begin{equation}
    \tilde{B}_t = \mathbf{B}_t + \gamma \cdot \Delta B_t, \quad \tilde{C}_t = \mathbf{C}_t + \gamma \cdot \Delta C_t,
\label{eq:modulate}
\end{equation}
where the scaling factor \(\gamma\) is initialized to a small value so that training begins near vanilla Mamba behavior, with the language influence gradually increasing during optimization. Substituting the modulated projections into the discretized recurrence (Eq.~\ref{eq:recurrent}), the language-conditioned dynamics become:
\begin{equation}
    h_t = \bar{A}_t \, h_{t-1} + \Delta_t \, \tilde{B}_t \, x_t, \quad y_t = \tilde{C}_t \, h_t.
\label{eq:lanmo_recurrent}
\end{equation}
Through \(\tilde{B}_t\), language priors reshape what information each patch writes into the hidden state, encouraging the accumulation of diagnosis-relevant evidence while attenuating irrelevant features. Through \(\tilde{C}_t\), they further control what context is read out at each step, ensuring that emitted representations emphasize diagnostically meaningful patterns. Consequently, the selective mechanism becomes diagnosis-aware without altering the SSM's temporal dynamics, enabling stable, linguistically guided aggregation of distributed pathological evidence.

\subsection{Local-Adaptive Vocabulary Retrieval Module}
\label{sec:vocab}
Prior VLM-MIL methods~\cite{vilamil,top,moc} typically rely on a fixed language vocabulary shared across all patches, ignoring the fact that different tissue regions demand different semantic priors. We instead design a local-adaptive vocabulary retrieval module that identifies task-relevant tissue concepts via a large language model (LLM) and dynamically assigns each patch fine-grained, tissue-specific language based on its local tissue identity. We now describe how the language embedding \(\mathbf{e}_t\) is constructed for each patch through this adaptive retrieval process. The central idea is to route each patch to semantically appropriate textual descriptors based on tissue identity confidence: when the tissue type can be reliably identified, fine-grained tissue-specific language is retrieved; otherwise, the module falls back to coarse-grained pathological descriptions to avoid introducing noisy priors.

\noindent \textbf{Task-specific concept set construction.}
For each downstream task, we prompt an LLM with the task definition and pathological background to generate the tissue types most diagnostically relevant to the target prediction, yielding a task-specific diagnostic concept set \(\mathcal{T} = \{T_1, \dots, T_M\}\), where each concept corresponds to a distinct tissue type. For each concept \(T_m\), we further prompt the LLM to generate a fine-grained language set \(\mathcal{L}_{m}\) containing textual descriptions that characterize the morphology, architectural patterns, and diagnostic implications of that tissue type. In parallel, we define a coarse-grained language set \(\mathcal{L}_{\mathrm{coarse}}\) that captures broader pathological categories and serves as a fallback when tissue identity is ambiguous.

\noindent \textbf{Confidence-gated concept routing.}
Each patch is routed to the most appropriate concept via CONCH~\cite{conch}. Let \(v_t\) denote the CONCH image embedding of patch \(x_t\) and \(q_m\) the CONCH text embedding of concept \(T_m\). We compute the cosine similarity \(a_{t,m} = \cos(v_t, q_m)\) and let \(m^{*} = \arg\max_{m}\, a_{t,m}\). The assignment is accepted only when the best similarity exceeds a threshold and the margin over the runner-up is sufficiently large:
\begin{equation}
    \label{eq:routing}
    a_{t,m^{*}} > \tau_{\mathrm{sim}}
    \qquad \text{and} \qquad
    a_{t,m^{*}} - \max_{m \neq m^{*}} a_{t,m} > \tau_{\mathrm{margin}}.
\end{equation}
The language source for patch \(x_t\) is then selected as
\begin{equation}
    \mathcal{L}(x_t) =
    \begin{cases}
        \mathcal{L}_{m^{*}}, & \text{if the assignment is accepted},\\
        \mathcal{L}_{\mathrm{coarse}}, & \text{otherwise}.
    \end{cases}
\end{equation}

\noindent \textbf{Language embedding construction.}
Given $\mathcal{L}(x_t)$, we encode each descriptor with the CONCH~\cite{conch} text encoder and aggregate the resulting embeddings to form the patch-level language representation $\mathbf{e}_t \in \mathbb{R}^{d}$, which is then fed into the low-rank modulation mechanism of Sec.~\ref{sec:KHiM-Mamba}. Through confidence-gated routing, $\mathbf{e}_t$ receives precise diagnostic language when tissue identity is clear and falls back to a safe semantic anchor otherwise, yielding modulation signals that are both semantically precise and robust across diverse tissue compositions compared with fixed-vocabulary approaches.

\newcommand{\thickhline}{\noalign{\hrule height 0.9pt}}

\section{Experiments}
We compare the proposed KHiM-Mamba against a broad spectrum of SoTA methods across 11 public datasets spanning cancer sub-typing, survival prediction. For cancer sub-typing, experiments are conducted on the BRACS~\cite{bracs} and TCGA-BRCA~\cite{tcga} datasets. For survival prediction, we evaluate on six TCGA~\cite{tcga} subsets, namely Bladder Urothelial Carcinoma (BLCA), Breast Invasive Carcinoma (BRCA), Kidney Renal Clear Cell Carcinoma (KIRC), Kidney Renal Papillary Cell Carcinoma (KIRP), Lung Adenocarcinoma (LUAD), and Stomach Adenocarcinoma (STAD). Beyond standard in-distribution evaluation, we further assess generalization under domain shift through cross-center adaptation experiments on the CAMELYON~\cite{camelyon16} benchmarks, where models are trained on CAMELYON16 and tested on CAMELYON17. Furthermore, to examine the performance under scarce data scenarios, we conduct few-shot experiments on TCGA-NSCLC~\cite{tcga} under 8-shot and 16-shot settings.

\subsection{Implementation Details}
All experiments are implemented in PyTorch 2.5.1. For WSI pre-processing, we employ the CLAM~\cite{clam} toolkit to segment each slide into non-overlapping $512 \times 512$ patches, whose features are then extracted using the CONCH~\cite{conch} vision encoder. The learning rate is set to $1\times10^{-5}$ for the BRACS, TCGA-BRCA, and CAMELYON datasets, and $1\times10^{-4}$ for the survival prediction task. For Local-Adaptive Vocabulary Retrieval Module in Sec.~\ref{sec:vocab}, we use GPT-5.6-sol as the base LLM (More ablations are shown in Tab.~\ref{ablation_retrieval}). We train KHiM-Mamba and all competing methods for 100 epochs with a batch size of 1 on a single NVIDIA RTX 3090 GPU. Scaling factor $\gamma$ in eq.~\ref{eq:modulate} is initialized to $1\times e^{-2}$. The routing thresholds $\tau_{sim}$ and $\tau_{margin}$ in eq.~\ref{eq:routing} are set to 0.25 and 0.03 respectively. More details including system prompts will be released with codes upon acceptance.

\begin{table*}[!t]
\centering
\caption{Comparison Results of Cancer Sub-typing Task using CONCH~\cite{conch} Features. $p$ denotes paired t-tests between our KHiM-Mamba and the second-best counterpart on the corresponding dataset.}
\resizebox{\textwidth}{!}{
\begin{tabular}{l|ccl|ccl|ccl}
\thickhline
\cellcolor{pink!30} &
\multicolumn{3}{c|}{\cellcolor{pink!30}\textbf{BRACS-7 (n=547, $p<0.05$)}} &
\multicolumn{3}{c|}{\cellcolor{pink!30}\textbf{BRACS-3 (n=547, $p<0.01$)}} &
\multicolumn{3}{c}{\cellcolor{pink!30}\textbf{TCGA-BRCA (n=952, $p<0.05$)}}  \\
\cellcolor{pink!30}\multirow{-2}{*}{\textbf{Method}} & \cellcolor{pink!30}\textbf{ACC.} \scriptsize($\pm$std) & \cellcolor{pink!30}\textbf{AUC} \scriptsize($\pm$std) & \cellcolor{pink!30}\textbf{{RANK}}
& \cellcolor{pink!30}\textbf{ACC.} \scriptsize($\pm$std) & \cellcolor{pink!30}\textbf{AUC} \scriptsize($\pm$std) & \cellcolor{pink!30}\textbf{{RANK}}
& \cellcolor{pink!30}\textbf{ACC.} \scriptsize($\pm$std) & \cellcolor{pink!30}\textbf{AUC} \scriptsize($\pm$std) & \cellcolor{pink!30}\textbf{{RANK}} \\
\thickhline
Mean Pooling & ${0.5378}_{\pm{0.0575}}$ & ${0.8341}_{\pm{0.0261}}$ & \textcolor{softred}{{\textbf{$\mathbf{15.5}$}}} & ${0.7746}_{\pm{0.0442}}$ & ${0.8769}_{\pm{0.0248}}$ & \textcolor{softred}{{\textbf{$\mathbf{15}$}}} & ${0.8970}_{\pm{0.0402}}$ & ${0.9204}_{\pm{0.0385}}$ & \textcolor{softred}{\textbf{$\mathbf{17}$}} \\

Max Pooling & ${0.5315}_{\pm{0.0426}}$ & ${0.8084}_{\pm{0.0588}}$ & \textcolor{softred}{{\textbf{$\mathbf{19.5}$}}} & ${0.7463}_{\pm{0.0424}}$ & ${0.8782}_{\pm{0.0203}}$ & \textcolor{softred}{{\textbf{$\mathbf{19.5}$}}} & ${0.8812}_{\pm{0.0399}}$ & ${0.9042}_{\pm{0.0449}}$ & \textcolor{softred}{\textbf{$\mathbf{22}$}} \\ \hline

ABMIL~\cite{abmil} \scriptsize \textcolor{gray}{[ICML'21]} & ${0.5378}_{\pm{0.0354}}$ & ${0.8268}_{\pm0.0280}$ & \textcolor{softred}{{\textbf{$\mathbf{16.5}$}}} & ${0.7571}_{\pm{0.0665}}$ & ${0.8730}_{\pm{0.0410}}$ & \textcolor{softred}{{\textbf{$\mathbf{19}$}}} & ${0.9028}_{\pm{0.0275}}$ & ${0.9354}_{\pm{0.0424}}$ & \textcolor{softgreen}{\textbf{$\mathbf{11}$}} \\

TransMIL~\cite{transmil_1} \scriptsize \textcolor{gray}{[NIPS'21]} & ${0.5149}_{\pm{0.0555}}$ & ${0.8190}_{\pm{0.0462}}$ & \textcolor{softred}{\textbf{$\mathbf{19}$}} & ${0.7632}_{\pm{0.0511}}$ & ${0.8953}_{\pm{0.0299}}$ & \textcolor{softgreen}{{\textbf{$\mathbf{9.5}$}}} & ${0.9006}_{\pm{0.0297}}$ & ${0.9302}_{\pm{0.0526}}$ & \textcolor{softred}{\textbf{$\mathbf{14}$}} \\

CLAM-SB~\cite{clam} \scriptsize \textcolor{gray}{[NAT.BME'20]} & ${0.5499}_{\pm{0.0521}}$ & ${0.8255}_{\pm{0.0308}}$ & \textcolor{softred}{{\textbf{$\mathbf{15.5}$}}} & ${0.7751}_{\pm{0.0505}}$ & ${0.8857}_{\pm{0.0369}}$ & \textcolor{softred}{{\textbf{$\mathbf{12}$}}} & ${0.9118}_{\pm{0.0368}}$ & ${0.9407}_{\pm{0.0418}}$ & \textcolor{softgreen}{{\textbf{$\mathbf{5}$}}} \\

CLAM-MB~\cite{clam} \scriptsize \textcolor{gray}{[NAT.BME'20]} & ${0.5673}_{\pm{0.0427}}$ & ${0.8554}_{\pm{0.0308}}$ & \textcolor{softgreen}{{\textbf{$\mathbf{6.5}$}}} & ${0.7782}_{\pm{0.0475}}$ & ${0.8937}_{\pm{0.0135}}$ & \textcolor{softgreen}{{\textbf{$\mathbf{7}$}}} & ${0.9064}_{\pm{0.0389}}$ & ${0.9366}_{\pm{0.0516}}$ & \textcolor{softgreen}{{\textbf{$\mathbf{8.5}$}}} \\

DSMIL~\cite{dsmil} \scriptsize \textcolor{gray}{[CVPR'21]} & ${0.5002}_{\pm{0.0709}}$ & ${0.8158}_{\pm{0.0344}}$ & \textcolor{softred}{{\textbf{$\mathbf{20.5}$}}} & ${0.7480}_{\pm{0.0344}}$ & ${0.8903}_{\pm{0.0281}}$ & \textcolor{softred}{{\textbf{$\mathbf{16.5}$}}} & ${0.8869}_{\pm{0.0463}}$ & ${0.9163}_{\pm{0.0568}}$ & \textcolor{softred}{{\textbf{$\mathbf{20}$}}} \\

ACMIL~\cite{acmil} \scriptsize \textcolor{gray}{[ECCV'24]} & ${0.5524}_{\pm{0.0506}}$ & ${0.8587}_{\pm{0.0297}}$ & \textcolor{softgreen}{\textbf{{$\mathbf{8}$}}} & ${0.7577}_{\pm{0.0456}}$ & ${0.8692}_{\pm{0.0250}}$ & \textcolor{softred}{{\textbf{$\mathbf{19}$}}} & ${0.9107}_{\pm{0.0369}}$ & ${0.9364}_{\pm{0.0423}}$ & \textcolor{softgreen}{{\textbf{$\mathbf{7}$}}} \\

ILRA-MIL~\cite{ilra} \scriptsize \textcolor{gray}{[ICLR'23]} & ${0.5029}_{\pm{0.0705}}$ & ${0.8243}_{\pm{0.0249}}$ & \textcolor{softred}{{\textbf{$\mathbf{19}$}}} & ${0.7594}_{\pm{0.0451}}$ & ${0.8810}_{\pm{0.0256}}$ & \textcolor{softred}{{\textbf{$\mathbf{16}$}}} & ${0.8953}_{\pm{0.0342}}$ & ${0.9252}_{\pm{0.0470}}$ & \textcolor{softred}{{\textbf{$\mathbf{17}$}}} \\

Dtfd-MIL~\cite{dtfdmil} \scriptsize \textcolor{gray}{[CVPR'22]} & ${0.5389}_{\pm{0.0652}}$ & ${0.8477}_{\pm{0.0591}}$ & \textcolor{softred}{{\textbf{$\mathbf{13.5}$}}} & ${0.7488}_{\pm{0.0396}}$ & ${0.8538}_{\pm{0.0382}}$ & \textcolor{softred}{{\textbf{$\mathbf{21}$}}} & ${0.8972}_{\pm{0.0488}}$ & ${0.9179}_{\pm{0.0382}}$ & \textcolor{softred}{{\textbf{$\mathbf{17}$}}} \\

MHIM-ABMI~\cite{MIHM-MIL} \scriptsize \textcolor{gray}{[IJCV'26]} & ${0.5405}_{\pm{0.0399}}$ & ${0.8328}_{\pm{0.0477}}$ & \textcolor{softred}{{\textbf{$\mathbf{15}$}}} & ${0.7678}_{\pm{0.0487}}$ & ${0.8832}_{\pm{0.0479}}$ & \textcolor{softred}{{\textbf{$\mathbf{14.5}$}}} & ${0.9033}_{\pm{0.0391}}$ & ${0.9345}_{\pm{0.0518}}$ & \textcolor{softred}{{\textbf{$\mathbf{11.5}$}}} \\ \hline

ViLa-MIL~\cite{vilamil}$\intercal$ \scriptsize \textcolor{gray}{[CVPR'24]} & ${0.5611}_{\pm{0.0567}}$ & ${0.8541}_{\pm{0.0749}}$ & \textcolor{softgreen}{{\textbf{$\mathbf{9.5}$}}} & ${0.7742}_{\pm{0.0392}}$ & ${0.8932}_{\pm{0.0488}}$ & \textcolor{softgreen}{{\textbf{$\mathbf{10.5}$}}} & ${0.9022}_{\pm{0.0395}}$ & ${0.9422}_{\pm{0.0676}}$ & \textcolor{softgreen}{{\textbf{$\mathbf{9}$}}} \\ 
FOCUS~\cite{focus}$\intercal$ \scriptsize \textcolor{gray}{[CVPR'25]} & ${0.5654}_{\pm{0.0468}}$ & $\uline{0.8644}_{\pm{0.0505}}$ & \textcolor{softgreen}{{\textbf{$\mathbf{4.5}$}}} & $\uline{0.7924}_{\pm{0.0672}}$ & ${0.8950}_{\pm{0.0648}}$ & \textcolor{softgreen}{{\textbf{$\mathbf{3.5}$}}} & ${0.9110}_{\pm{0.0785}}$ & ${0.9325}_{\pm{0.0612}}$ & \textcolor{softgreen}{{\textbf{$\mathbf{7.5}$}}} \\ 
TOP~\cite{top}$\intercal$ \scriptsize \textcolor{gray}{[NIPS'23]} & ${0.5631}_{\pm{0.0722}}$ & ${0.8573}_{\pm{0.0684}}$ & \textcolor{softgreen}{{\textbf{$\mathbf{7}$}}} & ${0.7794}_{\pm{0.0395}}$ & ${0.8930}_{\pm{0.0433}}$ & \textcolor{softgreen}{{\textbf{$\mathbf{7.5}$}}} & ${0.9047}_{\pm{0.0588}}$ & ${0.9391}_{\pm{0.0551}}$ & \textcolor{softgreen}{{\textbf{$\mathbf{8.5}$}}} \\ 
MOC~\cite{moc}$\intercal$ \scriptsize \textcolor{gray}{[MICCAI'25]} & ${0.5563}_{\pm{0.0675}}$ & ${0.8431}_{\pm{0.0499}}$ & \textcolor{softred}{{\textbf{$\mathbf{12}$}}} & ${0.7780}_{\pm{0.0617}}$ & ${0.8779}_{\pm{0.0682}}$ & \textcolor{softred}{{\textbf{$\mathbf{13}$}}} & ${0.8994}_{\pm{0.0467}}$ & ${0.9164}_{\pm{0.0259}}$ & \textcolor{softred}{{\textbf{$\mathbf{17}$}}} \\ \hline

S4MIL~\cite{S4MIL}$^{\dag}$ \scriptsize \textcolor{gray}{[MICCAI'23]} & ${\uline{0.5697}}_{\pm{0.0463}}$ & ${0.8557}_{\pm{0.0291}}$ & \textcolor{softgreen}{{\textbf{$\mathbf{4.5}$}}} & ${0.7839}_{\pm{0.0716}}$ & ${0.8963}_{\pm{0.0479}}$ & \textcolor{softgreen}{{\textbf{$\mathbf{3.5}$}}} & ${\uline{0.9133}}_{\pm{0.0276}}$ & ${0.9445}_{\pm{0.0377}}$ & \textcolor{softgreen}{{\textbf{$\mathbf{3}$}}} \\

MambaMIL~\cite{mambamil}$^{\dag}$ \scriptsize \textcolor{gray}{[MICCAI'24]} & ${0.5665}_{\pm{0.0703}}$ & ${0.8383}_{\pm{0.0336}}$ & \textcolor{softgreen}{{\textbf{$\mathbf{10}$}}} & ${0.7688}_{\pm{0.0296}}$ & ${0.8936}_{\pm{0.0270}}$ & \textcolor{softgreen}{{\textbf{$\mathbf{10.5}$}}} & ${0.9070}_{\pm{0.0403}}$ & ${\uline{0.9458}}_{\pm{0.0285}}$ & \textcolor{softgreen}{{\textbf{$\mathbf{4}$}}} \\

2DMamba~\cite{2dmamba}$^{\dag}$ \scriptsize \textcolor{gray}{[CVPR'25]} & ${0.5608}_{\pm{0.0491}}$ & ${0.8479}_{\pm{0.0405}}$ & \textcolor{softgreen}{{\textbf{$\mathbf{10.5}$}}} & ${{0.7858}}_{\pm{0.0475}}$ & ${\uline{0.8977}}_{\pm{0.0567}}$ & \textcolor{softgreen}{{\textbf{$\mathbf{2.5}$}}} & ${0.8957}_{\pm{0.0388}}$ & ${0.9347}_{\pm{0.0469}}$ & \textcolor{softred}{{\textbf{$\mathbf{14.5}$}}} \\

M3Mamba~\cite{m3mamba}$^{\dag}$$\ddag$ \scriptsize \textcolor{gray}{[CVPR'25]} & ${0.5692}_{\pm{0.0488}}$ & ${0.8581}_{\pm{0.0457}}$ & \textcolor{softgreen}{{\textbf{$\mathbf{4}$}}} & ${0.7544}_{\pm{0.0581}}$ & ${0.8875}_{\pm{0.0632}}$ & \textcolor{softred}{{\textbf{$\mathbf{16}$}}} & ${0.8857}_{\pm{0.0492}}$ & ${0.9147}_{\pm{0.0651}}$ & \textcolor{softred}{{\textbf{$\mathbf{21}$}}} \\

GMMamba~\cite{gmmamba}$^{\dag}$ \scriptsize \textcolor{gray}{[ICCV'25]} & ${\uline{0.5697}}_{\pm{0.0477}}$ & ${{0.8589}}_{\pm{0.0369}}$ & \textcolor{softgreen}{{\textbf{$\mathbf{2.5}$}}} & ${0.7774}_{\pm{0.0498}}$ & ${0.8945}_{\pm{0.0388}}$ & \textcolor{softgreen}{{\textbf{$\mathbf{7.5}$}}} & ${0.9069}_{\pm{0.0572}}$ & ${0.9441}_{\pm{0.0389}}$ & \textcolor{softgreen}{{\textbf{$\mathbf{5.5}$}}} \\ 

PAM~\cite{PAM}$^{\dag}$ \scriptsize \textcolor{gray}{[TMI'25]} & ${0.5682}_{\pm{0.0581}}$ & ${0.8542}_{\pm{0.0434}}$ & \textcolor{softgreen}{{\textbf{$\mathbf{6.5}$}}} & ${{0.7792}}_{\pm{0.0568}}$ & ${0.8904}_{\pm{0.0425}}$ & \textcolor{softgreen}{{\textbf{$\mathbf{8.5}$}}} & ${0.9066}_{\pm{0.0579}}$ & ${0.9249}_{\pm{0.0577}}$ & \textcolor{softred}{{\textbf{$\mathbf{12}$}}} \\ \thickhline

\rowcolor{violet!10}\textbf{KHiM-Mamba (Ours)}$\dag$$\intercal$ & ${\textbf{0.5874}}_{\pm{0.0857}}$ & ${\textbf{0.8746}}_{\pm{0.0459}}$ &  \textcolor{softgreen}{\textbf{$\mathbf{1}$}} &  ${\textbf{0.8165}}_{\pm{0.0398}}$ & ${\textbf{0.9074}}_{\pm{0.0469}}$ & \textcolor{softgreen}{\textbf{{$\mathbf{1}$}}} & ${\textbf{0.9330}}_{\pm{0.0427}}$ & ${\textbf{0.9559}}_{\pm{0.0433}}$ & \textcolor{softgreen}{\textbf{$\mathbf{1}$}} \\ \thickhline
\end{tabular}
}
\begin{threeparttable}
    \begin{tablenotes}
        \scriptsize 
        \item[] $\ddag$ M3Mamba~\cite{m3mamba} did not release their codes, results are from our reproduced version;
        \item[] $\dag$ denotes Mamba-based multiple instance learning methods. 
        \item[] $\intercal$ denotes multiple-instance learning methods that incorporate language priors.
    \end{tablenotes}
\end{threeparttable}
\label{tab:subtyping CONCH}
\vspace{-0.5cm}
\end{table*}

\subsection{Results in Cancer Sub-typing Tasks}
\noindent \textbf{Datasets \& Evaluation Metrics.} We evaluate on two benchmarks. The BRACS dataset~\cite{bracs} comprises 547 breast tumor WSIs annotated with three coarse categories (benign, atypical, and malignant) and seven fine-grained sub-types (Normal, Pathological Benign, Usual Ductal Hyperplasia, Flat Epithelial Atypia, Atypical Ductal Hyperplasia, Ductal Carcinoma in Situ, and Invasive Carcinoma). We report results under 10-fold Monte Carlo cross-validation with an 80\%/10\%/10\% train/val/test split. The TCGA-BRCA dataset~\cite{tcga} contains 952 breast WSIs divided into Invasive Ductal Carcinoma (IDC, 749 slides) and Invasive Lobular Carcinoma (ILC, 203 slides). Following GMMamba~\cite{gmmamba}, we randomly partition it into training, validation, and test sets at a 65:10:25 ratio. For both datasets, we report the mean Area Under the ROC Curve (AUC) with standard deviation (std).
\noindent \textbf{Results on BRACS.}
On the BRACS benchmark, KHiM-Mamba consistently ranks first under both the fine-grained 7-class setting and the coarse-grained 3-class setting, outperforming conventional MIL approaches, language-informed methods, and recent Mamba-based models. Specifically, on BRACS-7, KHiM-Mamba achieves an accuracy of 0.5874 and an AUC of 0.8746. The previous best accuracy is 0.5697, jointly achieved by S4MIL~\cite{S4MIL} and GMMamba~\cite{gmmamba}, while the previous best AUC is 0.8644, obtained by the language-informed FOCUS~\cite{focus}. KHiM-Mamba therefore improves the two metrics by 1.77 and 1.02 percentage points, respectively. These results demonstrate the effectiveness of incorporating localized linguistic knowledge into selective state-space modeling for challenging fine-grained subtype recognition. On BRACS-3, KHiM-Mamba similarly achieves the best accuracy and AUC of 0.8165 and 0.9074, respectively. Compared with the strongest previous results---an accuracy of 0.7924 from FOCUS~\cite{focus} and an AUC of 0.8977 from 2DMamba~\cite{2dmamba}---our method yields improvements of 2.41 and 0.97 percentage points ($p<0.05$). The consistent gains under both label granularities indicate that KHiM-Mamba can effectively capture discriminative morphological patterns for fine-grained classification while maintaining reliable performance when diagnostic categories are consolidated. Moreover, its improvements over existing language-informed methods suggest that locally adaptive language modulation is more effective than incorporating language priors through globally shared prompts or representations.

\noindent \textbf{Results on TCGA-BRCA.}
On TCGA-BRCA, KHiM-Mamba again ranks first, achieving an accuracy of 0.9330 and an AUC of 0.9559. It advances the previous best accuracy of 0.9133, obtained by S4MIL~\cite{S4MIL}, by 1.97 percentage points and the previous best AUC of 0.9458, obtained by MambaMIL~\cite{mambamil}, by 1.01 percentage points ($p<0.05$). These improvements are particularly notable because many competitive methods already exceed 0.90 in accuracy and 0.94 in AUC on this benchmark, leaving relatively limited room for further gains. The consistent superiority of KHiM-Mamba over classical attention-based MIL approaches, existing language-informed methods, and recent Mamba-family variants highlights the effectiveness of its language-modulated selective dynamics. Overall, achieving the top rank across all three settings and all six evaluation metrics demonstrates the robustness and generalizability of KHiM-Mamba across datasets with different subtype granularities and cohort characteristics.
\begin{table*}[!t]
    \centering
    \caption{Comparison Results of Survival Analysis Task on Six TCGA Subsets.}
    \resizebox{\textwidth}{!}{
    \begin{tabular}{l|cccccc|cl}
        \thickhline \rowcolor{pink!30} \textbf{Method}  & \textbf{BLCA \scriptsize (n=437)}   & \textbf{BRCA \scriptsize (n=1023)}   & \textbf{KIRC \scriptsize (n=498)} & \textbf{KIRP \scriptsize (n=261)} & \textbf{LUAD \scriptsize (n=455)} & \textbf{STAD \scriptsize (n=363)}   & \textbf{MEAN} & \textbf{RANK} \\
        \thickhline
        Mean-Pooling    & ${0.6758}_{\pm{0.0333}}$ & ${0.6274}_{\pm{0.0368}}$ & ${0.7287}_{\pm{0.0348}}$  & ${0.8179}_{\pm{0.0426}}$ & ${0.6334}_{\pm{0.0897}}$   & ${0.5917}_{\pm{0.0412}}$ & $0.6792$ & \textcolor{softred}{\textbf{$\mathbf{19}$}} \\
        Max-Pooling  & ${0.6417}_{\pm{0.0446}}$ & ${0.6095}_{\pm{0.1011}}$ & ${0.7316}_{\pm{0.0557}}$ & ${0.8164}_{\pm{0.0743}}$ & ${0.6245}_{\pm{0.0671}}$ & ${0.5554}_{\pm{0.0423}}$ & 0.6632 &   \textcolor{softred}{\textbf{$\mathbf{21}$}}\\\hline
        
        ABMIL~\cite{abmil} \scriptsize \textcolor{gray}{[ICML'21]}  & ${0.6604}_{\pm{0.0183}}$  & ${0.6621}_{\pm{0.0263}}$ & ${0.7185}_{\pm{0.0609}}$ & ${0.8201}_{\pm{0.0480}}$ & ${0.6433}_{\pm{0.0766}}$  & ${0.6053}_{\pm{0.0329}}$ & $0.6850$  &    \textcolor{softred}{\textbf{$\mathbf{17}$}}      \\

        TransMIL~\cite{transmil_1} \scriptsize \textcolor{gray}{[NIPS'21]}  & ${0.6705}_{\pm{0.0244}}$ & ${0.6592}_{\pm{0.0593}}$   & ${0.7174}_{\pm{0.0708}}$ & ${0.8357}_{\pm{0.0477}}$  & ${0.6185}_{\pm{0.1102}}$             & ${0.6597}_{\pm{0.0374}}$ & $0.6935$    &        \textcolor{softred}{\textbf{$\mathbf{11}$}} \\
        
        CLAM-SB~\cite{clam} \scriptsize \textcolor{gray}{[NAT.BME'20]}  & ${0.6711}_{\pm{0.0296}}$     & ${0.6650}_{\pm{0.0310}}$    & ${0.7323}_{\pm{0.0561}}$    & ${0.8216}_{\pm{0.0390}}$ & ${0.6387}_{\pm{0.0675}}$  & ${0.6154}_{\pm{0.0274}}$ &$0.6907$ &\textcolor{softred}{\textbf{$\mathbf{14}$}}\\

        CLAM-MB~\cite{clam} \scriptsize \textcolor{gray}{[NAT.BME'20]}  & ${0.6684}_{\pm{0.0275}}$     & ${0.6609}_{\pm{0.0476}}$    & ${0.7293}_{\pm{0.0583}}$    & ${0.8358}_{\pm{0.0642}}$ & ${0.6154}_{\pm{0.0779}}$  & ${0.5983}_{\pm{0.0506}}$ & $0.6848$& \textcolor{softred}{\textbf{$\mathbf{18}$}}\\

        DSMIL~\cite{dsmil}$^{\bot}$ \scriptsize \textcolor{gray}{[CVPR'21]}  & ${0.6751}_{\pm{0.0185}}$ & ${0.6828}_{\pm{0.0457}}$   & $\uline{0.7473}_{\pm{0.0478}}$ & ${0.8331}_{\pm{0.0626}}$  & ${0.6568}_{\pm{0.0901}}$ & ${0.6204}_{\pm{0.0587}}$  & $0.7026$ & \textcolor{softgreen}{\textbf{$\mathbf{7}$}}\\
        
        ACMIL~\cite{acmil} \scriptsize \textcolor{gray}{[ECCV'24]}  & ${0.6682}_{\pm{0.0379}}$ & ${0.6755}_{\pm{0.0322}}$   & ${0.7301}_{\pm{0.0584}}$ & ${0.8314}_{\pm{0.0492}}$  & ${0.6355}_{\pm{0.0914}}$  & ${0.6140}_{\pm{0.0456}}$ & $0.6925$ & \textcolor{softred}{\textbf{$\mathbf{13}$}}\\

        ILRA-MIL~\cite{ilra} \scriptsize \textcolor{gray}{[ICLR'23]}  & ${0.6802}_{\pm{0.0236}}$ & ${0.6705}_{\pm{0.0297}}$   & ${0.7335}_{\pm{0.0442}}$ & ${0.8368}_{\pm{0.0287}}$  & ${0.6417}_{\pm{0.0668}}$  & ${0.6451}_{\pm{0.0659}}$ & $0.7013$ & \textcolor{softgreen}{\textbf{$\mathbf8$}} \\
        
        Dtfd-MIL~\cite{dtfdmil} \scriptsize \textcolor{gray}{[CVPR'22]}  & ${0.6649}_{\pm{0.0279}}$ & ${0.6648}_{\pm{0.0387}}$   & ${0.7279}_{\pm{0.0458}}$ & ${0.8244}_{\pm{0.0577}}$  & ${0.6222}_{\pm{0.0477}}$  & ${0.6355}_{\pm{0.0728}}$ & $0.6900$ & \textcolor{softred}{\textbf{$\mathbf{15}$}} \\

        MHIM-MIL~\cite{MIHM-MIL} \scriptsize \textcolor{gray}{[IJCV'26]}  & ${0.6752}_{\pm{0.0348}}$ & ${0.6705}_{\pm{0.0299}}$   & ${0.7145}_{\pm{0.0699}}$ & ${0.8259}_{\pm{0.0491}}$  & ${0.6470}_{\pm{0.0833}}$  & ${0.6440}_{\pm{0.0562}}$ & $0.6961$ & \textcolor{softgreen}{\textbf{$\mathbf{10}$}}\\ \thickhline

        WSI-GCN~\cite{wsi-gcn}$^{\ddag}$ \scriptsize \textcolor{gray}{[MICCAI'21]}  & ${0.6729}_{\pm{0.0417}}$ & ${0.6578}_{\pm{0.0588}}$   & ${0.7050}_{\pm{0.0233}}$ & ${0.8378}_{\pm{0.0620}}$  & ${0.6467}_{\pm{0.0969}}$  & ${0.6372}_{\pm{0.0588}}$ & $0.6929$ & \textcolor{softred}{\textbf{$\mathbf{12}$}}\\ 

        ED-GNN~\cite{ed-gnn}$^{\ddag}$  \scriptsize \textcolor{gray}{[MICCAI'24]} & ${0.6822}_{\pm{0.0459}}$ & ${0.6675}_{\pm{0.0490}}$   & ${0.6978}_{\pm{0.0391}}$ & ${0.7946}_{\pm{0.0285}}$  & ${0.6333}_{\pm{0.0629}}$  & ${0.6490}_{\pm{0.0599}}$ & $0.6874$ & \textcolor{softred}{\textbf{$\mathbf{16}$}}\\\thickhline

        VLSA~\cite{VLSA}$\intercal$  \scriptsize \textcolor{gray}{[ICLR'25]} & ${0.6176}_{\pm{0.0215}}$ & ${0.6652}_{\pm{0.0570}}$   & ${0.7157}_{\pm{0.0579}}$ & ${0.7822}_{\pm{0.0647}}$  & ${0.6370}_{\pm{0.0270}}$  & ${0.6475}_{\pm{0.0298}}$ & $0.6766$ & \textcolor{softred}{\textbf{$\mathbf{20}$}}\\\thickhline

        S4-MIL~\cite{S4MIL}$\dag$ \scriptsize \textcolor{gray}{[MICCAI'23]}  & ${0.6811}_{\pm{0.0314}}$ & ${0.6686}_{\pm{0.0609}}$   & ${0.7378}_{\pm{0.0341}}$ & ${0.8296}_{\pm{0.0544}}$  & ${0.6445}_{\pm{0.1075}}$  & ${0.6579}_{\pm{0.0693}}$ & $0.7033$ &\textcolor{softgreen}{\textbf{$\mathbf6$}} \\

        MambaMIL~\cite{mambamil}$\dag$ \scriptsize \textcolor{gray}{[MICCAI'24]}  & ${0.6837}_{\pm{0.0465}}$ & ${0.6841}_{\pm{0.0258}}$   & ${0.7388}_{\pm{0.0436}}$ & ${0.8257}_{\pm{0.0369}}$  & ${0.6509}_{\pm{0.0594}}$  & ${0.6632}_{\pm{0.0459}}$ & $0.7077$ & \textcolor{softgreen}{\textbf{$\mathbf3$}} \\

        2DMamba~\cite{2dmamba}$\dag$ \scriptsize \textcolor{gray}{[CVPR'25]}  & $\uline{0.6902}_{\pm{0.0384}}$ & $\uline{0.6977}_{\pm{0.0567}}$   & ${0.7290}_{\pm{0.0391}}$ & $\uline{0.8465}_{\pm{0.0277}}$  & $\uline{0.6644}_{\pm{0.0498}}$  & $\uline{0.6700}_{\pm{0.0389}}$ & $\uline{0.7163}$ & \textcolor{softgreen}{\textbf{$\mathbf2$}} \\

        M3Mamba~\cite{m3mamba}$\dag$ \scriptsize \textcolor{gray}{[CVPR'25]}  & ${0.6675}_{\pm{0.0392}}$ & ${0.6738}_{\pm{0.0388}}$   & ${0.7279}_{\pm{0.0577}}$ & ${0.8144}_{\pm{0.0690}}$  & ${0.6549}_{\pm{0.0485}}$  & ${0.6605}_{\pm{0.0720}}$ & $0.6998$ & \textcolor{softgreen}{\textbf{$\mathbf9$}} \\

        GMMamba~\cite{gmmamba}$\dag$ \scriptsize \textcolor{gray}{[ICCV'25]}  & ${0.6795}_{\pm{0.0458}}$ & ${0.6877}_{\pm{0.0382}}$   & ${0.7274}_{\pm{0.0492}}$ & ${0.8233}_{\pm{0.0491}}$  & ${0.6571}_{\pm{0.0538}}$  & ${0.6687}_{\pm{0.0597}}$ & $0.7073$ & \textcolor{softgreen}{\textbf{$\mathbf4$}} \\ 

        PAM~\cite{PAM}$\dag$ \scriptsize \textcolor{gray}{[TMI'25]}  & ${0.6749}_{\pm{0.0392}}$ & ${0.6878}_{\pm{0.0261}}$   & ${0.7355}_{\pm{0.0544}}$ & ${0.8198}_{\pm{0.0374}}$  & ${0.6600}_{\pm{0.0475}}$  & ${0.6639}_{\pm{0.0488}}$ & $0.7070$ & \textcolor{softgreen}{\textbf{$\mathbf5$}} \\ 
        
        \thickhline
        \rowcolor{violet!10} \textbf{KHiM-Mamba (Ours)}$\dag$$\intercal$ & ${\mathbf{0.7243}}_{\pm{0.0488}}$ & $\mathbf{0.7175}_{\pm{0.0384}}$   & $\mathbf{0.7579}_{\pm{0.0592}}$ & $\mathbf{0.8577}_{\pm{0.0489}}$  & $\mathbf{0.6858}_{\pm{0.0376}}$  & $\mathbf{0.6929}_{\pm{0.0457}}$ & $\mathbf{0.7394}$ & \textcolor{softgreen}{\textbf{$\mathbf1$}} \\
    \thickhline
    \end{tabular}
    }
    \begin{threeparttable}
        \begin{tablenotes}
            \scriptsize \item[] $\dag$ denotes Mamba-based multiple instance learning methods; 
            \item[] $\ddag$ Graph-based multiple instance learning methods \textbf{specifically designed for survival analysis}
            \item[] $\intercal$ denotes multiple-instance learning methods that incorporate language priors; among prior VLM-MIL methods, only VLSA~\cite{VLSA} supports regression-based survival analysis.
        \end{tablenotes}
    \end{threeparttable}
    \label{tab:survival}
    \vspace{-0.5cm}
\end{table*}
\begin{figure*}[pos=t]
    \centering
    \includegraphics[width=\textwidth]{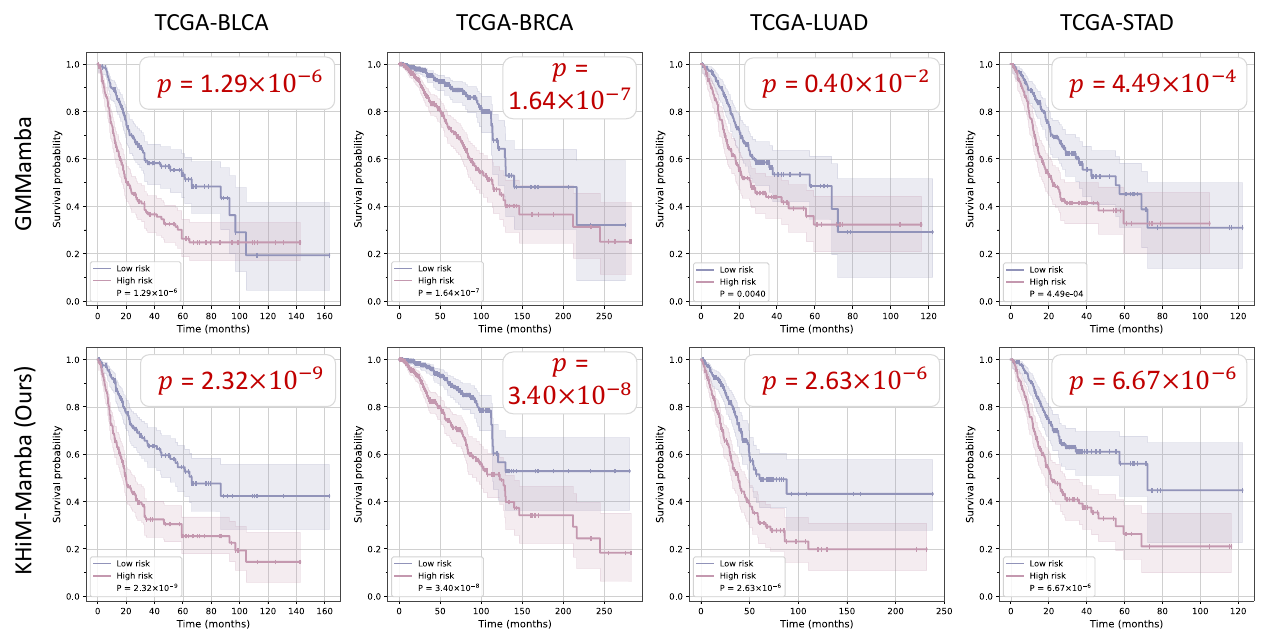}
    \caption{Kaplan–Meier survival curves of second-best method GMMamba~\cite{gmmamba} and our proposed KHiM-Mamba.} \label{fig:km_curve}
    \vspace{-0.5cm}
\end{figure*}
\subsection{Results in Survival Outcome Prediction Tasks}
\noindent \textbf{Datasets \& Evaluation Metrics.}
We evaluate on six TCGA subsets (BLCA, BRCA, KIRC, KIRP, LUAD, and STAD), using overall survival length as the ground-truth label following~\cite{mambamil}. A 5-fold cross-validation strategy with a 4:1 train/validation split is adopted to mitigate the impact of data partitioning. We report the mean cross-validated Concordance Index (C-Index) with std.
\noindent \textbf{Results on Six TCGA Subsets.}
As shown in Table~\ref{tab:survival}, KHiM-Mamba achieves the highest mean C-Index of 0.7394 across the six TCGA cohorts, with concordance indices of 0.7243, 0.7175, 0.7579, 0.8577, 0.6858, and 0.6929 on BLCA, BRCA, KIRC, KIRP, LUAD, and STAD, respectively. This represents a 2.31\% improvement over the strongest prior methods, 2DMamba (0.7163), and consistently surpasses other recent Mamba-based methods such as MambaMIL (0.7077), GMMamba (0.7073), PAM (0.7070), and M3Mamba (0.6998). The consistent gains over these architecturally diverse SSM variants indicate that the benefit of our language-modulated selective dynamics is complementary to, and extends beyond, purely structural enhancements to the state-space backbone.

To further evaluate risk stratification quality, we present Kaplan–Meier survival curves in Fig.~\ref{fig:km_curve}. Compared with the second-best method GMMamba~\cite{gmmamba}, KHiM-Mamba yields consistently lower log-rank $p$-values across all four datasets, indicating sharper separation between predicted high- and low-risk groups. The improvement is especially pronounced on TCGA-LUAD, where the 
$p$-value drops from $4.00\times10^{-2}$ (GMMamba) to $2.63\times10^{-6}$ (KHiM-Mamba), reflecting a substantially more discriminative risk stratification. These results suggest that language-modulated slide encoding dynamics enable the model to capture survival-relevant evidence more effectively, producing risk scores that better align with actual patient outcomes.
\begin{table*}[!t]
    \centering
    \caption{Domain Generalization Performance on CAMELYON16 to CAMELYON17~\cite{camelyon16}.}
    \resizebox{\textwidth}{!}{
    \begin{tabular}{l|ccccc|cl}
        \thickhline \rowcolor{pink!30} \textbf{Method}  & \textbf{CENTER 1}   & \textbf{CENTER 2}   & \textbf{CENTER 3} & \textbf{CENTER 4} & \textbf{CENTER 5} & \textbf{MEAN}   & \textbf{RANK} \\
        \thickhline
        Mean-Pooling    & ${0.6760}_{\pm{0.1237}}$ & ${0.3780}_{\pm{0.0183}}$ & ${0.7850}_{\pm{0.0838}}$ & ${0.6758}_{\pm{0.1063}}$ & ${0.6647}_{\pm{0.0301}}$ & $0.6359$ & \textcolor{softred}{\textbf{$\mathbf{16}$}}\\
        Max-Pooling & ${0.9210}_{\pm{0.0164}}$ & ${0.7780}_{\pm{0.1454}}$ & ${0.9510}_{\pm{0.0094}}$ & ${0.9333}_{\pm{0.0112}}$ & ${0.8330}_{\pm{0.0200}}$ & $0.8833$ & \textcolor{softgreen}{\textbf{$\mathbf4$}}\\  \thickhline
        ABMIL~\cite{abmil} \scriptsize \textcolor{gray}{[ICML'21]}    & ${0.9120}_{\pm{0.0248}}$ & ${0.7450}_{\pm{0.1344}}$ & ${0.9480}_{\pm{0.0098}}$ & ${0.9323}_{\pm{0.0197}}$ & ${0.8460}_{\pm{0.0918}}$ & $0.8767$ & \textcolor{softgreen}{\textbf{$\mathbf7$}} \\
        TransMIL~\cite{transmil_1} \scriptsize \textcolor{gray}{[NIPS'21]}    & $\uline{0.9230}_{\pm{0.0245}}$ & $\uline{0.8450}_{\pm{0.0880}}$ & ${0.9440}_{\pm{0.0191}}$ & ${0.9333}_{\pm{0.0176}}$ & ${0.8330}_{\pm{0.0424}}$ & $\uline{0.8957}$ & \textcolor{softgreen}{\textbf{$\mathbf2$}} \\
        CLAM~\cite{clam} \scriptsize \textcolor{gray}{[NAT.BME'20]}    & ${0.9020}_{\pm{0.0252}}$ & ${0.7170}_{\pm{0.1254}}$ & ${0.9440}_{\pm{0.0080}}$ & ${0.9253}_{\pm{0.0164}}$ & ${0.8080}_{\pm{0.0583}}$ & $0.8593$ & \textcolor{softred}{\textbf{$\mathbf{15}$}}\\
        ACMIL~\cite{acmil} \scriptsize \textcolor{gray}{[ECCV'24]}    & ${0.9020}_{\pm{0.0312}}$ & ${0.7460}_{\pm{0.1247}}$ & ${0.9490}_{\pm{0.0137}}$ & ${0.9273}_{\pm{0.0185}}$ & ${0.7930}_{\pm{0.0880}}$ & $0.8635$ & \textcolor{softred}{\textbf{$\mathbf{14}$}}\\
        ILRA-MIL~\cite{ilra} \scriptsize \textcolor{gray}{[ICLR'23]}    & ${0.9070}_{\pm{0.0398}}$ & ${0.7680}_{\pm{0.1408}}$ & ${0.9350}_{\pm{0.0191}}$ & ${0.9242}_{\pm{0.0244}}$ & ${0.8100}_{\pm{0.0986}}$ & $0.8688$ & \textcolor{softred}{\textbf{$\mathbf{12}$}}\\ \thickhline
        
        ViLa-MIL~\cite{vilamil}$\intercal$ \scriptsize \textcolor{gray}{[CVPR'24]}    & ${0.9070}_{\pm{0.0743}}$ & ${0.7720}_{\pm{0.0695}}$ & ${0.9420}_{\pm{0.0484}}$ & ${0.9330}_{\pm{0.0575}}$ & ${0.8170}_{\pm{0.0549}}$ & ${0.8742}$ & \textcolor{softred}{\textbf{$\mathbf{9}$}} \\   
        FOCUS~\cite{focus}$\intercal$ \scriptsize \textcolor{gray}{[CVPR'25]}    & ${0.9140}_{\pm{0.0579}}$ & ${0.7740}_{\pm{0.0688}}$ & ${0.9430}_{\pm{0.0482}}$ & ${0.9290}_{\pm{0.0550}}$ & ${0.8280}_{\pm{0.0398}}$ & ${0.8776}$ & \textcolor{softgreen}{\textbf{$\mathbf{6}$}} \\
        TOP~\cite{top}$\intercal$ \scriptsize \textcolor{gray}{[NIPS'23]}    & ${0.9130}_{\pm{0.0576}}$ & ${0.7680}_{\pm{0.0499}}$ & ${0.9360}_{\pm{0.0642}}$ & ${0.9240}_{\pm{0.0833}}$ & ${0.8180}_{\pm{0.0755}}$ & ${0.8718}$ & \textcolor{softred}{\textbf{$\mathbf{10}$}} \\ 
        MOC~\cite{moc}$\intercal$ \scriptsize \textcolor{gray}{[MICCAI'25]}    & ${0.9150}_{\pm{0.0398}}$ & ${0.7630}_{\pm{0.0469}}$ & ${0.9270}_{\pm{0.0574}}$ & ${0.9272}_{\pm{0.0648}}$ & ${0.8190}_{\pm{0.0555}}$ & ${0.8702}$ & \textcolor{softred}{\textbf{$\mathbf{11}$}} \\ \thickhline
        
        S4MIL~\cite{S4MIL}$\dag$ \scriptsize \textcolor{gray}{[MICCAI'23]}    & ${0.9210}_{\pm{0.0217}}$ & ${0.8060}_{\pm{0.1038}}$ & ${0.9430}_{\pm{0.0168}}$ & $\uline{0.9354}_{\pm{0.0164}}$ & ${0.8320}_{\pm{0.0462}}$ & $0.8875$ & \textcolor{softgreen}{\textbf{$\mathbf3$}} \\ 
        MambaMIL~\cite{mambamil}$\dag$ \scriptsize \textcolor{gray}{[MICCAI'24]}    & ${0.9200}_{\pm{0.0358}}$ & ${0.7510}_{\pm{0.1797}}$ & ${0.9390}_{\pm{0.0181}}$ & ${0.9253}_{\pm{0.0213}}$ & ${0.8400}_{\pm{0.0755}}$ & $0.8751$ & \textcolor{softgreen}{\textbf{$\mathbf8$}} \\ 
        GMMamba~\cite{gmmamba}$\dag$ \scriptsize \textcolor{gray}{[ICCV'25]}    & ${0.9070}_{\pm{0.0452}}$ & ${0.7190}_{\pm{0.1550}}$ & ${0.9300}_{\pm{0.0195}}$ & ${0.9303}_{\pm{0.0178}}$ & ${0.8350}_{\pm{0.0819}}$ & $0.8643$ & \textcolor{softred}{\textbf{$\mathbf{13}$}}\\ 
        PAM~\cite{PAM}$\dag$ \scriptsize \textcolor{gray}{[TMI'25]}    & ${0.9170}_{\pm{0.0241}}$ & ${0.7500}_{\pm{0.1579}}$ & $\uline{0.9540}_{\pm{0.0092}}$ & ${0.9283}_{\pm{0.0071}}$ & $\uline{0.8600}_{\pm{0.0377}}$ & $0.8819$ & \textcolor{softgreen}{\textbf{$\mathbf5$}}\\ \thickhline
        \rowcolor{violet!10} \textbf{KHiM-Mamba (Ours)}$\dag$$\intercal$ & $\textbf{0.9477}_{\pm{0.0418}}$ & $\textbf{0.8654}_{\pm{0.0279}}$ & $\textbf{0.9560}_{\pm{0.0722}}$ & $\textbf{0.9457}_{\pm{0.0275}}$ & $\textbf{0.8772}_{\pm{0.0588}}$ & $\textbf{0.9184}$ & \textcolor{softgreen}{\textbf{$\mathbf1$}} \\ 
    \thickhline
    \end{tabular}
    }
    \begin{threeparttable}
        \begin{tablenotes}
            \scriptsize \item[] $\dag$ denotes Mamba-based multiple instance learning methods;
            \item[] $\intercal$ denotes multiple-instance learning methods that incorporate language priors
        \end{tablenotes}
    \end{threeparttable}
    \label{tab:camelyon}
    \vspace{-0.7cm}
\end{table*}

\subsection{Cross-center Generalization Evaluation}
For domain generalization evaluation, all models are trained on CAMELYON16 and directly evaluated on the five independent centers of CAMELYON17 without any target-domain adaptation. Since these centers were prepared and digitized under different staining protocols and scanners, this cross-center protocol serves as a direct stress test of the diagnostically irrelevant appearance variation illustrated in Fig.~\ref{fig:intro_1}b. As shown in Table~\ref{tab:camelyon}, KHiM-Mamba achieves the best performance on all five target centers, with AUC scores of 0.9477, 0.8654, 0.9560, 0.9457, and 0.8772, respectively. The resulting mean AUC is 0.9184, which is the highest among all compared methods. The strongest previous method is TransMIL with a mean AUC of 0.8957, followed by S4MIL (0.8875), PAM (0.8819), ABMIL (0.8767), and MambaMIL (0.8751). Relative to TransMIL, KHiM-Mamba improves the mean AUC by 2.27\%, indicating stronger generalization under cross-center distribution shifts. The superiority of KHiM-Mamba is also consistent across individual centers. Compared with the best competing result on each center, our method achieves gains of 2.47\% on Center 1, 2.04 on Center 2, 0.20\% on Center 3, 1.24\% on Center 4, and 1.72\% on Center 5. In particular, the improvement on Center 2 is notable, where several recent MIL and Mamba-based methods exhibit relatively lower transfer performance. Overall, these results show that KHiM-Mamba maintains stronger and more stable performance when transferred from Camelyon16 to the multi-center Camelyon17 benchmark.
\subsection{Performance under Few-Shot Scenarios}
To evaluate data efficiency, we conduct few-shot experiments on TCGA-NSCLC under 8-shot and 16-shot settings, where only a handful of labeled slides per class are available for training. Results are reported in Table~\ref{tab:few-shot}. Under the 8-shot setting, KHiM-Mamba achieves 81.45\% accuracy and 93.48\% AUC, surpassing the most SoTA method MOC by 4.35\% and 2.97\%, respectively. Conventional MIL methods such as CLAM-SB degrade sharply in this regime (59.96\% accuracy), while existing vision--language approaches like ViLa-MIL and CoOp fail to consistently outperform purely visual methods, suggesting that naively incorporating language is insufficient under severe data scarcity. When the shot count increases to 16, KHiM-Mamba continues to lead with 84.77\% accuracy and 92.58\% AUC. Averaged over both settings, our method attains 83.11\% accuracy and 93.03\% AUC with relatively low variance across runs, confirming that language-modulated selective dynamics serve as a strong inductive bias that substantially alleviates the dependence on annotation.
\begin{table}[!t]
    \centering
    \caption{\label{tab:few-shot}Few-shot performance on the TCGA-NSCLC~\cite{tcga} dataset.}
    \setlength{\tabcolsep}{1pt}
    \resizebox{0.48\textwidth}{!}{
    \begin{tabular}{l|cc|cc|cc}
        \toprule
        \cellcolor{pink!30} &
        \multicolumn{2}{c|}{\cellcolor{pink!30}\textbf{8-SHOT}} &
        \multicolumn{2}{c|}{\cellcolor{pink!30}\textbf{16-SHOT}} &
        \multicolumn{2}{c|}{\cellcolor{pink!30}\textbf{MEAN}} \\
        
        \cellcolor{pink!30}\multirow{-2}{*}{\textbf{Method}} & \cellcolor{pink!30}\textbf{ACC.} 
        & \cellcolor{pink!30}\textbf{AUC} 
        & \cellcolor{pink!30}\textbf{ACC.}  
        & \cellcolor{pink!30}\textbf{AUC}
        & \cellcolor{pink!30}\textbf{ACC.} & \cellcolor{pink!30}\textbf{AUC} \\
        \midrule
        CLAM-SB~\cite{clam} & ${59.96}_{\pm{9.44}}$ & ${73.68}_{\pm{6.72}}$ & ${78.52}_{\pm{4.16}}$ & ${87.03}_{\pm{3.78}}$ & $69.11$ & $80.36$\\
        CLAM-MB~\cite{clam} & ${66.63}_{\pm{2.48}}$ & ${79.69}_{\pm{6.34}}$ & ${73.76}_{\pm{3.88}}$ & ${88.01}_{\pm{2.99}}$ & $70.20$ & $83.85$\\
        TransMIL~\cite{transmil_1} & ${75.82}_{\pm{7.44}}$ & ${83.69}_{\pm{7.05}}$ & ${73.88}_{\pm{2.58}}$ & ${88.53}_{\pm{3.17}}$ & $74.85$& $86.11$\\
        ViLa-MIL~\cite{vilamil} & ${73.51}_{\pm{9.22}}$ & ${84.20}_{\pm{7.09}}$ & ${71.28}_{\pm{3.56}}$ & ${88.30}_{\pm{4.12}}$ & $72.40$& $86.25$ \\
        CoOp~\cite{coop} & ${69.48}_{\pm{5.93}}$ & ${80.11}_{\pm{5.91}}$ & ${71.81}_{\pm{4.35}}$ & ${82.10}_{\pm{3.72}}$ & $70.65$& $81.11$\\
        TOP~\cite{top} & ${71.02}_{\pm{7.52}}$ & ${79.43}_{\pm{8.30}}$ & $\uline{79.26}_{\pm{6.72}}$ & ${86.42}_{\pm{4.55}}$ & $\uline{75.14}$& $82.93$\\
        MOC~\cite{moc} & $\uline{77.10}_{\pm{3.80}}$ & $\uline{90.51}_{\pm{1.74}}$ & ${72.40}_{\pm{2.46}}$ & $\uline{89.95}_{\pm{2.77}}$ & $74.75$& $\uline{90.23}$\\ \thickhline
        \cellcolor{violet!10}\textbf{KHiM-Mamba} & \cellcolor{violet!10}$\textbf{81.45}_{\pm{3.75}}$ & \cellcolor{violet!10}$\textbf{93.48}_{\pm{3.79}}$ & \cellcolor{violet!10}$\textbf{84.77}_{\pm{4.69}}$ & \cellcolor{violet!10}$\textbf{92.58}_{\pm{3.75}}$ & \cellcolor{violet!10}$\textbf{83.11}$& \cellcolor{violet!10}$\textbf{93.03}$\\ \thickhline
    \end{tabular}}
    \vspace{-0.2cm}
\end{table}

\subsection{Ablation Study}
In this section, we conduct a series of ablation experiments to dissect the contribution of each design choice in KHiM-Mamba. Unless otherwise specified, all ablation studies are using AUC as the evaluation metric.
\begin{table}[!t]
    \begin{centering}
        \caption{\label{ablation}Ablations on our core Knowledge-aware Hidden-state Modulation (KHiM) module upon cancer sub-typing datasets, \textit{i.e.}, BRACS-7, TCGA-BRCA, and CAMELYON16 tumor region localization (patch-level classification) task.}
        \setlength{\tabcolsep}{1pt}
        \resizebox{0.48\textwidth}{!}{
        \begin{tabular}{l|c|c|c|c|c}
            \thickhline \rowcolor{pink!30} \textbf{Settings} & \textbf{$\mathbf{\Delta B}$} & \textbf{$\mathbf{\Delta C}$} & \textbf{BRACS-7} & \textbf{TCGA-BRCA} & \textbf{CAMELYON16} \tabularnewline \thickhline 
            {Original SSM} & \ding{55} & \ding{55} & $0.8277_{\pm0.0542}$ & $0.9046_{\pm0.0372}$ & $0.8573_{\pm0.0549}$ \tabularnewline 
            
            \texttt{Version 1} & \ding{52} & \ding{55} & $\uline{0.8333}_{\pm0.0439}$ & $\uline{0.9067}_{\pm0.0538}$ & ${0.8745}_{\pm0.0576}$ \tabularnewline
            
            \texttt{Version 2} & \ding{55} & \ding{52} & $0.8325_{\pm0.0643}$ & $0.8976_{\pm0.0422}$ & $0.8620_{\pm0.0628}$ \tabularnewline \thickhline 
            
            \rowcolor{violet!10} \textbf{Ours} & \ding{52} & \ding{52} & $\mathbf{0.8746}_{\pm0.0459}$ & $\mathbf{0.9559}_{\pm0.0433}$ & $\mathbf{0.9075}_{\pm0.0482}$ \tabularnewline \thickhline
        \end{tabular}}
        \vspace{-0.5cm}
    \end{centering}
\end{table}

\noindent \textbf{Effect of KHiM Components.}
KHiM-Mamba injects linguistic priors into the selective SSM by modulating the input projection matrices associated with $B_t$ and $C_t$ through learned low-rank residuals $\Delta B$ and $\Delta C$. To assess their individual and joint contributions, we compare four variants: (i) \texttt{Original SSM}, which uses a standard Mamba backbone without language conditioning; (ii) \texttt{Version 1}, which applies only $\Delta B$; (iii) \texttt{Version 2}, which applies only $\Delta C$; and (iv) the full \textbf{KHiM-Mamba}, which jointly applies both residuals. The results on two cancer subtyping datasets and the CAMELYON16 tumor region localization are in Table~\ref{ablation}.

Applying $\Delta B$ alone consistently improves performance over the original SSM, increasing the average score across the three tasks from 0.8632 to 0.8715. In contrast, applying only $\Delta C$ yields an average score of 0.8640, which is comparable to the baseline. This observation is intuitive: modulating $B_t$ affects what information is written into the hidden state, whereas modulating $C_t$ adjusts how the stored information is retrieved without directly influencing its encoding. When $\Delta B$ and $\Delta C$ are jointly activated, KHiM-Mamba achieves an average score of 0.9127, outperforming the baseline by 4.95 percentage points. Consistent improvements are observed across all three tasks, including BRACS-7 (0.8277 $\rightarrow$ 0.8746), TCGA-BRCA (0.9046 $\rightarrow$ 0.9559), and CAMELYON16 tumor region segmentation (0.8573 $\rightarrow$ 0.9075). These results demonstrate that $\Delta B$ and $\Delta C$ are complementary: language-guided modulation of $B_t$ enables the model to selectively encode task-relevant morphological patterns, while modulation of $C_t$ facilitates the effective retrieval of the corresponding discriminative features.

\noindent \textbf{Effect of Local-Adaptive Vocabulary Retrieval Module.}
We investigate how the quality and source of language priors affect language-guided modulation by comparing our local-adaptive vocabulary retrieval module with two alternative vocabulary settings: (i) raw class labels and (ii) the fixed prompt sets adopted by VilaMIL~\cite{vilamil} and TOP~\cite{top}. We further evaluate vocabularies generated by different LLMs, including GPT-5.6-sol, Gemini-3-pro, LLaMA, and Qwen-3.8-max, to assess the robustness of our retrieval strategy to the choice of language model.

As shown in Tab.~\ref{ablation_retrieval}, using only class labels yields the lowest mean AUC of 0.8939, suggesting that such coarse semantic supervision is insufficient for fine-grained feature modulation. Replacing class labels with manually designed and globally shared prompts from VilaMIL and TOP improves the mean AUC to 0.9090 and 0.9083, respectively, confirming that richer linguistic descriptions provide more effective guidance for slide representation learning.

In contrast, our local-adaptive retrieval strategy consistently achieves stronger performance across all tested LLMs, with mean AUCs ranging from 0.9084 to 0.9153. Among them, {GPT-5.6-sol} delivers the best overall results, achieving the highest mean AUC of {0.9153}, together with the best performance on both {BRACS-7} ({0.8746}) and {TCGA-BRCA} ({0.9559}). {Gemini-3-pro} also performs competitively, reaching a mean AUC of {0.9125}. Notably, the \textit{open-source} model {LLaMA-3.1}~\cite{touvron2023llama} and Qwen-3.8-max~\cite{qwen3} remain highly competitive with the proprietary models: {Qwen-3.8-max} attains a mean AUC of {0.9115} and the second-best result on {BRACS-7} ({0.8720}), while {LLaMA} achieves a mean AUC of {0.9084}, comparable to the strongest fixed-prompt baselines. Overall, the best configuration improves over class-label supervision by 2.14 percentage points and surpasses the strongest fixed-prompt baseline by 0.63 percentage points. These results demonstrate that the proposed module can effectively leverage both proprietary and open-source LLMs to produce informative local-adaptive vocabularies, showing strong effectiveness and generalizability.

\begin{table}[!t]
    \begin{centering}
        \caption{\label{ablation_retrieval}Ablations on our Local-Adaptive Vocabulary Retrieval module upon cancer sub-typing datasets.}
        \setlength{\tabcolsep}{1pt}
        \resizebox{0.42\textwidth}{!}{
        \begin{tabular}{l|c|c|c}
            \thickhline \rowcolor{pink!30} \textbf{Vocabulary} &  \textbf{BRACS-7} & \textbf{TCGA-BRCA} & \textbf{MEAN} \tabularnewline \thickhline
            Class-label & ${0.8466}_{\pm0.0451}$ & ${0.9411}_{\pm0.0675}$ & $0.8939$ \\
            VilaMIL~\cite{vilamil} Prompt & ${{0.8671}}_{\pm0.0420}$ & ${0.9509}_{\pm0.0483}$ & $0.9090$ \\
            TOP~\cite{top} Prompt & ${0.8644}_{\pm0.0387}$ & ${{0.9521}}_{\pm0.0576}$ & $0.9083$ \\ \thickhline
            \rowcolor{violet!10} {Ours (GPT-5.6-sol)} & $\textbf{0.8746}_{\pm0.0459}$ & $\textbf{0.9559}_{\pm0.0433}$ & $\textbf{0.9153}$ \\
            \rowcolor{violet!10} {Ours (Gemini-3-pro)} & ${0.8710}_{\pm0.0279}$ & $\uline{0.9540}_{\pm0.0533}$ & $\uline{0.9125}$ \\
            \rowcolor{violet!10} {Ours (LLaMA-3.1)} & ${0.8648}_{\pm0.0338}$ & ${0.9520}_{\pm0.0572}$ & ${0.9084}$ \\
            \rowcolor{violet!10} {Ours (Qwen-3.8-max)} & $\uline{0.8720}_{\pm0.0337}$ & ${0.9510}_{\pm0.0478}$ & ${0.9115}$
            \tabularnewline \thickhline
        \end{tabular}} \par
    \end{centering}
\end{table}

\subsection{Visualization Analysis of KHiM-Mamba:} 
\begin{figure}[pos=t]
    \centering
    \includegraphics[width=0.49\textwidth]{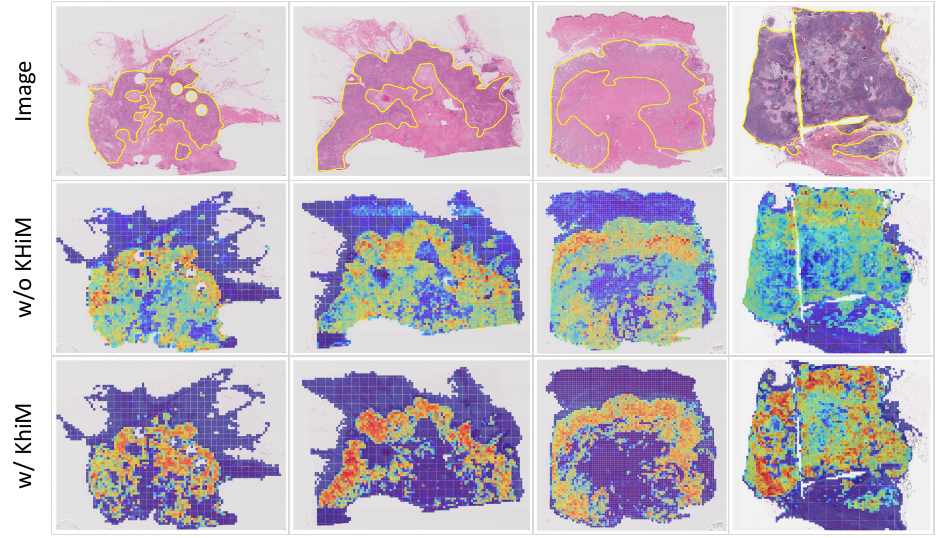}
    \vspace{-0.3cm}
    \caption{Visualization of patch attention map using patch features with and without Knowledge-aware Hidden-state Modulation (KHiM). Major tumor areas are circled in yellow.} \label{fig:vis}
    \vspace{-0.5cm}
\end{figure}

To intuitively understand how language modulation reshapes patch representations, we visualize patch-level attention heatmaps before and after the KHiM module in Fig.~\ref{fig:vis}. Before modulation, attention maps exhibit diffuse, noisy activations scattered across both tumor and non-tumor areas, indicating that purely visual features struggle to identify diagnostically relevant tissue. After modulation, attention becomes markedly more concentrated: high-attention responses align closely with annotated tumor regions (yellow contours), while non-tumor areas are consistently suppressed. This demonstrates that language-modulated selective dynamics effectively steer the model's focus toward pathologically meaningful regions, filtering out irrelevant background and yielding more discriminative representation. Moreover, the improvement in tumor region localization on the CAMELYON16 dataset (see Tab.~\ref{ablation}) quantitatively supports the same conclusion.
\begin{table}[!t]
    \centering
    \caption{\label{tab:efficiency}
    Efficacy evaluation with different numbers of patches per slide (\textit{i.e.}, 1,000 and 5,000).
Parameter counts (Params.), inference time, and throughput (Thro.) are reported in thousands (K), milliseconds (ms), and thousands of patches per second (K patches/s), respectively.}
    \setlength{\tabcolsep}{2pt}
    \resizebox{0.48\textwidth}{!}{
    \begin{tabular}{l|c|cc|cc}
        \toprule
        \cellcolor{pink!30} &
        \cellcolor{pink!30} &
        \multicolumn{2}{c|}{\cellcolor{pink!30}\textbf{n=1,000}} &
        \multicolumn{2}{c}{\cellcolor{pink!30}\textbf{n=5,000}} \\
        
        \cellcolor{pink!30}\multirow{-2}{*}{\textbf{Method}} &
        \cellcolor{pink!30}\multirow{-2}{*}{
            \shortstack{\textbf{Params.}}} &
        \cellcolor{pink!30}\shortstack{\textbf{Time}} &
        \cellcolor{pink!30}\shortstack{\textbf{Thro.}} &
        \cellcolor{pink!30}\shortstack{\textbf{Time}} &
        \cellcolor{pink!30}\shortstack{\textbf{Thro.}} \\
        \midrule
        
        Mean-Pooling                   & $525.8$K  & $0.25$ms  & $4.02$K & $0.53$ms  & $1.90$K \\
        Max-Pooling                    & $525.8$K  & $0.19$ms  & $5.22$K & $0.51$ms  & $1.97$K \\
        CLAM~\cite{clam}\scriptsize \textcolor{gray}{[NAT.BME'20]}          & $790.8$K  & $0.82$ms  & $1.22$K & $0.94$ms  & $1.07$K \\
        ABMIL~\cite{abmil}\scriptsize \textcolor{gray}{[ICML'21]}        & $591.6$K  & $0.71$ms  & $1.41$K & $0.80$ms  & $1.25$K \\
        DSMIL~\cite{dsmil}\scriptsize \textcolor{gray}{[CVPR'21]}        & $872.7$K  & $1.48$ms  & $0.68$K & $1.95$ms  & $0.51$K \\
        Dtfd-MIL~\cite{dtfdmil} \scriptsize \textcolor{gray}{[TMI'25]}         & $789.3$K  & $9.81$ms  & $0.10$K & $13.86$ms & $0.07$K \\
        TransMIL~\cite{transmil_1}\scriptsize \textcolor{gray}{[NIPS'21]}  & $2672.2$K & $9.47$ms  & $0.11$K & $11.40$ms & $0.09$K \\ 
        MHIM-MIL~\cite{MIHM-MIL}\scriptsize \textcolor{gray}{[IJCV'26]}  & $5344.3$K & $10.51$ms  & $0.10$K & $12.10$ms & $0.08$K \\
        \thickhline
        PAM~\cite{PAM}\scriptsize \textcolor{gray}{[TMI'25]}            & $2290.4$K & $3.09$ms  & $0.32$K & $6.47$ms  & $0.16$K \\
        MambaMIL~\cite{mambamil}\scriptsize \textcolor{gray}{[MICCAI'24]}  & $2410.2$K & $2.87$ms  & $0.35$K & $4.96$ms  & $0.20$K \\
        2DMamba~\cite{2dmamba}\scriptsize \textcolor{gray}{[CVPR'25]}
                                  & $2288.4$K & $2.46$ms  & $0.41$K & $6.44$ms  & $0.16$K \\ 
        \thickhline
        \cellcolor{violet!10}\textbf{KHiM-Mamba (ours)} &
        \cellcolor{violet!10}$2216.7$K &
        \cellcolor{violet!10}$2.50$ms &
        \cellcolor{violet!10}$0.41$K &
        \cellcolor{violet!10}$5.79$ms &
        \cellcolor{violet!10}$0.17$K \\
        \thickhline
    \end{tabular}}
    \vspace{-0.2cm}
\end{table}
\subsection{Efficacy Analysis}
We further evaluate the computational efficiency of KHiM-Mamba under different sequence lengths by sampling 1,000 and 5,000 patches per slide. As shown in Table~\ref{tab:efficiency}, KHiM-Mamba contains 2.22M trainable parameters, fewer than all compared Mamba-based MIL methods, including PAM (2.29M), MambaMIL (2.41M), and 2DMamba (2.29M). With 1,000 patches, our method requires only 2.50,ms for inference and achieves a throughput of 0.41K, matching 2DMamba while being faster than PAM and MambaMIL. When the sequence length increases to 5,000 patches, KHiM-Mamba maintains an inference time of 5.79,ms, outperforming PAM and 2DMamba by 10.5\% and 10.1\%, respectively. Although the proposed knowledge-aware modulation introduces language-conditioned operations into the selective state-space layer, its low-rank formulation incurs only modest computational overhead and preserves the favorable efficiency of Mamba-based architectures. Together with the performance gains reported in Tables~1--4, these results demonstrate that KHiM-Mamba achieves a favorable balance between predictive effectiveness, model size, and computational efficiency for large-scale WSI analysis.
\section{Conclusion}
In this work, we introduced \textbf{K}nowledge-aware \textbf{Hi}dden-state \textbf{M}odulation (KHiM-Mamba) that addresses a fundamental limitation of existing Mamba-based WSI analysis: purely vision-driven state transitions can accumulate diagnostically irrelevant evidence and dilute critical cues over long patch sequences. Rather than employing language only as an external component like previous VLM-MIL methods, KHiM-Mamba directly incorporates language-derived knowledge into the selective state-space mechanism, thereby reshaping the internal slide encoding dynamics. Specifically, the proposed write-in and read-out modulation schemes regulate what visual evidence is accumulated in the evolving hidden state and what contextual information is retrieved to enrich patch representations. Moreover, our local-adaptive vocabulary retrieval module dynamically assigns fine-grained, tissue-specific semantic descriptions to individual patches, providing precise and task-adaptive knowledge priors. Extensive experiments on 11 public benchmarks across four evaluation settings demonstrate that KHiM-Mamba consistently outperforms state-of-the-art methods in cancer subtyping and survival prediction, while exhibiting strong generalization under multi-center domain adaptation and few-shot learning. These results highlight the potential of explicitly regulating hidden-state evolution with diagnostic knowledge and provide a new perspective for designing knowledge-guided state-space models for computational pathology.

\bibliographystyle{cas-model2-names}
\bibliography{tmi}

\end{document}